\documentclass[prl,twocolumn,tightenlines,superscriptaddress]{revtex4-1}
\usepackage{bm,dcolumn,amsmath,graphicx,amsfonts,amssymb}

\usepackage{hyperref} 
\usepackage{xcolor}
\definecolor{cite}{rgb}{0.,0.,0.85}   
\hypersetup{colorlinks,linkcolor={cite},citecolor={cite},urlcolor={cite}}

\newcommand{\braket}[1]{\ensuremath{\langle #1\rangle}}	

\renewcommand{\v}[1]{\ensuremath{\boldsymbol{#1}}}		

\def\d{\ensuremath{{\rm d}}}

\def\h{\ensuremath{\hbar}} 

\renewcommand{\a}{\ensuremath{\alpha}}

\newcommand{\s}{\ensuremath{\sigma}} 
\renewcommand{\t}{\ensuremath{\tau}}

\def\CT{\ensuremath{{\cal T}}}

\def\CL{\ensuremath{{\cal L}}}

\newcommand{\un}[1]{\ensuremath{\,{\rm{#1}}}} 

\newcommand{\be}{\begin{equation}}
\newcommand{\ee}{\end{equation}}

\renewcommand{\subsection}[1]{\vspace{0.5pt}\paragraph*{\textbf{\textit{\small{#1---}}}}}
\renewcommand{\subsubsection}[1]{\paragraph*{{\textit{\small{#1---}}}}}

\def\SYRTE{SYRTE, Observatoire de~Paris, Universit\'e~PSL, CNRS, Sorbonne~Universit\'e, LNE, 61 avenue de~l'Observatoire, 75014 Paris, France}
\def\NPL{National Physical Laboratory, Hampton Road, Teddington TW11 0LW, United Kingdom}
\def\PTB{Physikalisch-Technische Bundesanstalt, Bundesallee 100, 38116 Braunschweig, Germany}
\def\LPL{Laboratoire de Physique des Lasers, Universit\'e Paris 13, Sorbonne Paris Cit\'e, CNRS, 99 Avenue Jean-Baptiste Cl\'ement, 93430 Villetaneuse, France}
\def\RENATER{R\'eseau National de t\'el\'ecommunications pour la Technologie, l'Enseignement et la Recherche, 23--25 Rue Daviel, 75013 Paris, France}
\def\ROA{Secci\'on de Hora, Real Instituto y Observatorio de la Armada, San Fernando, Spain}
\def\BIPM{Bureau International des Poids et Mesures, BIPM, Pavillon de Breteuil, 92312 S\`evres, France}
\def\JILA{JILA, National Institute of Standards and Technology and University of Colorado, Department of Physics, University of Colorado, Boulder, CO 80309, USA}
\def\UQ{School of Mathematics and Physics, University of Queensland, Brisbane, QLD 4072, Australia}

\begin{document} 

\title{Search for transient variations of the fine structure constant\\ and dark matter using fiber-linked optical atomic clocks}

\author{B.~M.~Roberts}\email[]{b.roberts@uq.edu.au}\altaffiliation[Present address:~]{\UQ}\affiliation{\SYRTE}
\author{P.~Delva}\affiliation{\SYRTE}
\author{A.~Al-Masoudi}\affiliation{\PTB}
\author{A.~Amy-Klein}\affiliation{\LPL}
\author{C.~B{\ae}rentsen}\affiliation{\SYRTE}
\author{C.~F.~A.~Baynham}\affiliation{\NPL}
\author{E.~Benkler}\affiliation{\PTB}
\author{S.~Bilicki}\affiliation{\SYRTE}
\author{S.~Bize}\affiliation{\SYRTE}
\author{W.~Bowden}\affiliation{\NPL}
\author{J.~Calvert}\affiliation{\SYRTE}
\author{V.~Cambier}\affiliation{\SYRTE}
\author{E.~Cantin}\affiliation{\SYRTE}\affiliation{\LPL}
\author{E.~A.~Curtis}\affiliation{\NPL}
\author{S.~D\"orscher}\affiliation{\PTB}
\author{M.~Favier}\affiliation{\SYRTE}
\author{F.~Frank}\affiliation{\SYRTE}
\author{P.~Gill}\affiliation{\NPL}
\author{R.~M.~Godun}\affiliation{\NPL}
\author{G.~Grosche}\affiliation{\PTB}
\author{C.~Guo}\affiliation{\SYRTE}
\author{A.~Hees}\affiliation{\SYRTE}
\author{I.~R.~Hill}\affiliation{\NPL}
\author{R.~Hobson}\affiliation{\NPL}
\author{N.~Huntemann}\affiliation{\PTB}
\author{J.~Kronj\"ager}\affiliation{\NPL}
\author{S.~Koke}\affiliation{\PTB}
\author{A.~Kuhl}\affiliation{\PTB}
\author{R.~Lange}\affiliation{\PTB}
\author{T.~Legero}\affiliation{\PTB}
\author{B.~Lipphardt}\affiliation{\PTB}
\author{C.~Lisdat}\affiliation{\PTB}
\author{J.~Lodewyck}\affiliation{\SYRTE}
\author{O.~Lopez}\affiliation{\LPL}
\author{H.~S.~Margolis}\affiliation{\NPL}
\author{H.~\'Alvarez-Mart\'inez}\affiliation{\SYRTE}\affiliation{\ROA}
\author{F.~Meynadier}\affiliation{\SYRTE}\affiliation{\BIPM}
\author{F.~Ozimek}\affiliation{\NPL}
\author{E.~Peik}\affiliation{\PTB}
\author{P.-E.~Pottie}\affiliation{\SYRTE}
\author{N.~Quintin}\affiliation{\RENATER}
\author{C.~Sanner}\altaffiliation[Present address:~]{\JILA}\affiliation{\PTB}
\author{L.~De Sarlo}\affiliation{\SYRTE} 
\author{M.~Schioppo}\affiliation{\NPL}
\author{R.~Schwarz}\affiliation{\PTB}
\author{A.~Silva}\affiliation{\NPL}
\author{U.~Sterr}\affiliation{\PTB}
\author{Chr.~Tamm}\affiliation{\PTB}
\author{R.~Le~Targat}\affiliation{\SYRTE}
\author{P.~Tuckey}\affiliation{\SYRTE}
\author{G.~Vallet}\affiliation{\SYRTE}
\author{T.~Waterholter}\affiliation{\PTB}
\author{D.~Xu}\affiliation{\SYRTE}
\author{P.~Wolf}\email[]{peter.wolf@obspm.fr}\affiliation{\SYRTE}

\date{\today}

\begin{abstract}
We search for transient variations of the fine structure constant using data from a European network of fiber-linked optical atomic clocks.
By searching for coherent variations in the recorded clock frequency comparisons across the network, we significantly improve the constraints on transient variations of the fine structure constant.
For example, we constrain the variation to $|\delta\a/\a|<5\times10^{-17}$ for transients of duration $10^3$\,s.
This analysis also presents a possibility to search for dark matter, the mysterious substance hypothesised to explain galaxy dynamics and other astrophysical phenomena that is thought to dominate the matter density of the universe.
At the current sensitivity level, we find no evidence for dark matter in the form of topological defects (or, more generally, any macroscopic objects), and we thus place constraints on certain potential couplings between the dark matter and standard model particles, substantially improving upon the existing constraints, particularly for large ($\gtrsim10^4\,$km) objects.
\end{abstract} 

\maketitle

\subsection{Introduction}
The nature of dark matter is one of the most important outstanding problems in physics today.
Despite composing the majority of the matter in the universe, evidence for dark matter particles in direct detection experiments remains elusive~\cite{Liu2017}.
So far, much of the focus has been on weakly-interacting massive particles (WIMPs) with masses 
equivalent to $\gtrsim$\,GeV; the lack of evidence for their existence, however, is contributing to an increase in interest for more varied candidate models~\cite{Bertone2018}.

One possibility is that dark matter is composed of ultralight boson fields (masses~$\ll1\un{eV}$).
Such fields may form classical oscillating fields that can be coherent on certain time scales \cite{Preskill1983,Dine1983}.
If the fields have specific self-interactions, they may also form stable macroscopic objects such as topological defects \cite{Kibble1980,Vilenkin1985}.
If the fields have non-gravitational interactions with standard model fields,  encounters between such objects and precision measurement devices may induce observable transient signatures in recorded data as Earth moves through the galactic dark matter halo~\cite{DereviankoDM2014}.

Here, we consider topological defect dark matter objects that have quadratic scalar interactions with standard model particles.
Such interactions lead to the effective rescaling of certain fundamental constants, which can shift atomic energy levels and transition frequencies (see, e.g., the recent review in Ref.~\cite{AtomicReview2017}).
Searches for transient frequency variations can then be performed by monitoring atomic clocks, which work by referencing the frequency of an external oscillator (e.g.,\ a laser) to that of an atomic transition.

We note that ultralight dark matter can also cause long-term drifts \cite{Olive2002,StadnikDMalpha2015,GodNisJon14,HunLipTam14,McGrew2018a} and local oscillations \cite{Arvanitaki2014,Tilburg2015,Hees2016,Hees2018,DereviankoVULF2016,Geraci2018} of fundamental constants.
Dark matter with other couplings can also be sought with atomic clocks~\cite{Wolf2018a} and networks of  other precision measurement devices, such as magnetometers~\cite{Pospelov2013,Alonso2018}; such searches are complementary to those considered in this work.
While we specifically consider quadratic couplings, the analysis applies equally for linear couplings (for the correspondence see, e.g.,\ Ref.~\cite{Hees2018}), though these are more tightly constrained~\cite{Olive2008}.

With only a single measurement device, it is impossible to distinguish a transient frequency variation caused by a variation in fundamental constants from one caused by terrestrial sources.
With a distributed network, however, the time-delays between signals appearing across network nodes must be consistent with the passing of a galactic-speed transient.
On the time-scales considered in this work ($>$\,$60\,$s), this will manifest as a simultaneous signal visible in all data streams.
In addition, having a network with multiple different clock types helps to discriminate against false positives.
Different clock types will respond differently to effective changes in fundamental constants, with the relative sensitivities being a prediction of the theory~\cite{FlambaumCJP2009}.

In this work, we use data from a European network of fiber-linked optical atomic clocks to search for evidence of transient variations in clock frequencies.
Our analysis has allowed us to substantially improve constraints on transient variations of the fine structure constant, $\a$,
particularly for time scales above $\sim$\,$10^2$\,s, where the long-term stability of the atomic clock comparisons in this network offers the largest advantage over existing experiments.
For example, at  $\sim$\,$10^3$\,s, we constrain the transient variation to $|\delta\a/\a|<5\times10^{-17}$.
We consider only the variation of $\a$ since we employ optical clocks, which are only sensitive to this parameter~\cite{FlambaumCJP2009}.

This analysis can be interpreted in terms of a search for  dark matter in the form of topological defects, and we find no evidence for such objects at the current sensitivity level.
Assuming the defects make up the majority of the dark matter in the galaxy, we then place constraints on their possible couplings with standard model fields.
Our results substantially improve upon the existing limits, particularly for large defects ($\gtrsim$\,$10^4\un{km}$).

\subsection{Transient variations of constants}

Transient effects, in general, are associated with two distinct time scales.
Firstly, there is the duration of each transient effect, which we denote as $\t_{\rm int}$.
Secondly, there is the average time between consecutive transients, which we denote as $\CT$.
Due to better statistics, a more precise measurement (or a more stringent constraint) can be made for effects with longer transient durations.
However, this requires good long-term measurement stability (i.e., no drifts) in order to track the signal over time.
This is one benefit of laboratory clock-clock comparisons, which have excellent long-term frequency stability.
Constraints for the time between transients are limited by the observation time.

Since the observable of an atomic clock is its frequency, for a comparison of two clocks with frequencies $\nu_A$ and $\nu_B$, we define the ratio $y_{AB}\equiv\nu_A/\nu_B$.
The fractional variation in this ratio caused by a variation in $\a$ occurring during the sampling period $\t$
for a pair of clocks both located at position $\v{r}$ at time $t$ is 
\be \label{eq:dw}
\frac{\delta y}{y_{AB}}(\v{r},t) = \frac{1}{\tau}\int_{t-\t}^{t} K_{AB} \frac{\delta \a (\v{r},t')}{\a}\,\d t',
\ee
where $K_{AB}$ quantifies the sensitivity of the frequency ratio ${y_{AB}}$ to the variation in $\a$ \cite{Dzuba1999,*DzuFlaWebPRL1999}.
This factor depends both on the atomic species and the transition considered.

The clock output, driven by the external oscillator, is referenced to the frequency of the probed atomic transition for time-scales larger than the servo loop constant, $\t_{\rm servo}$.
In writing Eq.~(\ref{eq:dw}), we have assumed that the effective sampling interval is larger than the servo time: $\t>\t_{\rm servo}$.
The experiment will still have sensitivity to variations below this time-scale, and it is possible to extend the analysis by taking into account the clock and laser responses below the servo time.
Here, we focus only on the region where $\tau_{\rm int}>\t_{\rm servo}$, where optical clocks are the most efficient and the main advantage of the optical clock comparisons is realised.

Assuming the transient variation follows a Gaussian profile,
i.e.,\ $\delta\a(t) = \delta\a_0\exp(-(t-t_0)^2/\tau_{\rm int}^2)$,
Eq.~(\ref{eq:dw}) can be evaluated simply.
The maximum $\delta\a$-induced perturbation is
\be \label{eq:dw2}
\frac{\delta {y}_0}{{y_{AB}}} = 
\begin{cases}
 K_{AB} \dfrac{\delta\a_0}{\a}  \dfrac{\sqrt{\pi}\,\tau_{\rm int}}{\t}   & {\rm for}~~ \tau_{\rm int}\ll\t\\
 \\[-0.2cm]
K_{AB} \dfrac{\delta\a_0}{\a}   & {\rm for}~~ \tau_{\rm int}\gg\t,
\end{cases}
\ee
where $\delta\a_0$ is the maximum amplitude of the transient variation $\delta\a(t)$, which occurs at time $t_0$.
The result changes only slightly for other profiles; 
e.g., for a rectangular (top-hat) profile the $\sqrt{\pi}$ factor is absent.

From this, one can constrain the possible values for $\delta\a_0$ by monitoring ratios of atomic clock frequencies.
In the simplest case, the maximum allowed value for $\delta\a_0$ for a given $\t_{\rm int}$ is set by the maximum observed $\delta{y}_0$ at the same time scale.
An experiment with much greater sensitivity can be performed using a network of clocks, 
provided their instabilities are comparable, by searching for variations in the frequency ratios that are coherent across the entire network, and are consistent with a transient variation of $\a$ (given the known $K$ coefficients).

\subsection{Data and analysis}

\begin{figure}[t]
\centering
\includegraphics[width=0.35\textwidth]{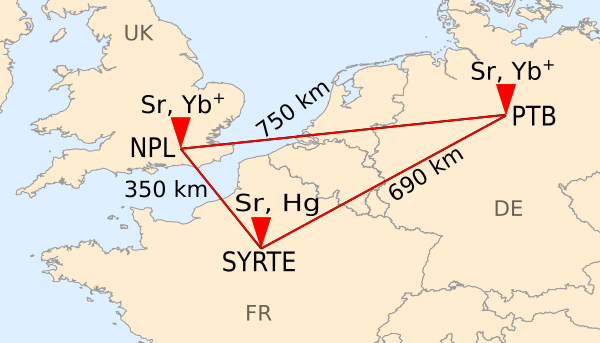}
\caption{European fiber-linked optical clock network.
The relevant lengths are the linear distances between laboratories, not the length of the actual optical fiber links.
The links use forward/backward light reflections to actively cancel signal variations coming from within the link~\cite{Lisdat2016}.
Therefore, the effect of variation of constants on the links themselves will not affect the results on time-scales longer than that of the round trip time.
The typical light reflection time is $10^{-3}\un{s}$, much shorter than the $\sim$\,$10^2$--$10^4\un{s}$ transients studied here.
}
\label{fig:map}
\end{figure}

We analyse data from a European network of fiber-linked optical atomic clocks based on Sr, Hg, and Yb$^+$ atoms, located in France, Germany,
and the United Kingdom, see Fig.~\ref{fig:map}.
The data was taken over a period of just over 40 days during May--June 2017.
The clocks' operation are described in Refs.~\cite{King2012,LeTargat2013,Hill2016,Lisdat2016,Grebing2016,Tyumenev2016,Sanner2019}.
The same fiber links have been used previously for fundamental physics tests, e.g.,\ in 
Ref.~\cite{Delva2017}. 
Due to the use of fiber links to perform the comparisons, the measurement stability is limited only by the instability of the clocks themselves, with negligible contributions from the fiber-based optical frequency transfer for the timescales $>60$\,s considered here~\cite{Lisdat2016}.

The base sampling interval of the data is $1\,$s.
However, we average the data from each clock stream up to the largest servo-time of the considered clocks, in order to be consistent with the assumption in Eq.~(\ref{eq:dw}).
The maximum servo times are those of the Yb$^+$ clocks ($\t_{\rm servo}^{\rm max}\sim$\,60\,s), so the effective sampling period is taken to be $\t=60\,$s. 
For averaging periods larger than this, the noise of all the clock pairs is essentially white frequency noise, with frequency instability scaling as $1/\sqrt{\t}$.
At $10^2$\,s averaging, the fractional frequency instability approaches $10^{-16}$ for the Sr-Yb$^+$ comparison at PTB,
and a few times $10^{-16}$ for the other local comparisons; more details are given in the Supplemental Material.
The relevant $K$ factors are $6.01$, $-0.75$, and $6.76$, for the Sr-Yb$^+$, Sr-Hg, and Hg-Yb$^+$ comparisons, respectively~\cite{Angstmann2004,DzubaKrel2008}.

If the source of the variation in $\a$ is galactic, we can expect it to move relative to Earth with galactic speeds, $v_g$\,$\sim$\,$300\un{km}\un{s}^{-1}$ (set, e.g.,\ by the motion of Earth through the galactic frame of rest).
If we assume that the relative velocity distribution for the transients is described by the standard halo model (as for dark matter, see, e.g.,\ Ref.~\cite{Freese2013}),  more than 99\% of the transients would move relative to Earth with $v\gtrsim75\un{km/s}$ \cite{GPSDM2017}.
In the condition that $\t_{\rm int}\gg L/v_g$, where $L$ is the distance between clocks, we can treat all the clocks in the network as being co-located, in that they will be affected simultaneously.
Since the longest distance in 750\un{km} as shown in Fig.~\ref{fig:map},
this condition is easily satisfied for the $\t_{\rm int}\gtrsim60\un{s}$ time-scales considered here.

We use a maximum-likelihood method similar to the approach developed in Ref.~\cite{GPSDM2018} to search for transient frequency variations across the network. 
The details of the method are given in the Supplemental Material.
In short, we define a likelihood function that quantifies how consistent the data covering a given time window is with a possible transient variation in $\a$.
We considered only time periods when at least two independent clock pairs (four clocks) were actively taking data, 
so that each clock appears only once in the combined data streams.
This eliminates cross-correlations between clock pairs, which would complicate the analysis.

We also define a detection threshold for the likelihood, above which there can be no false-positives with 99\% confidence.
Here, a false-positive is defined as any time the likelihood surpasses the threshold due to purely random noise processes.
Any time the likelihood is greater than the threshold can be investigated as a potential event.
No such instances were found using the considered data set, allowing us to place constraints on the $\a$ variation.

For each time window throughout the total observation time, we calculate the best-fit value for $\delta\a_0$ (denoted $\delta\a_0^{\rm bf}$) that maximises the likelihood for each relevant value of the possible interaction duration $\t_{\rm int}$.
The method also provides an estimate of the uncertainty, $\Delta\a_0$, in this best-fit value. 
Constraints can be placed by finding the largest best-fit $\delta\a_0^{\rm bf}$ that appears throughout the span of the data as a function of $\t_{\rm int}$, taking the uncertainty into account for the confidence level:
$| {\delta\a_0}| < |{\delta\a_0^{\rm bf}}|_{\rm max} +\Delta\a_0$; see the Supplemental Material for more details.

\begin{figure}
\includegraphics[width=0.48\textwidth]{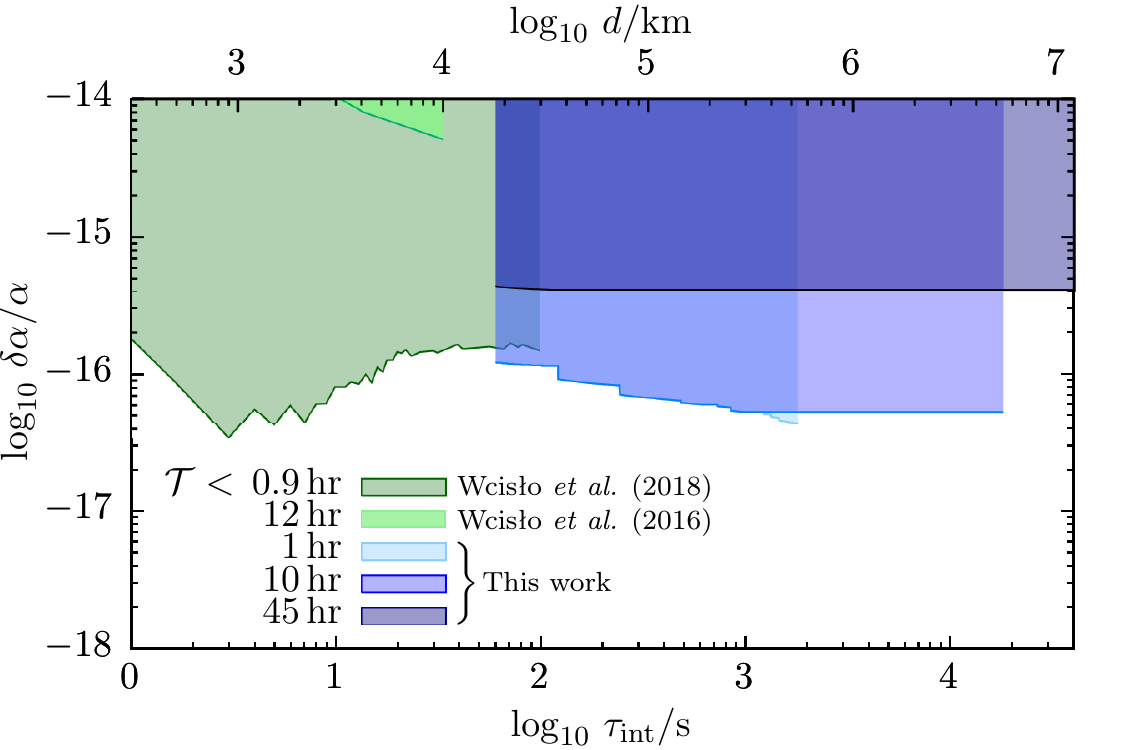}
\caption{Constraints on the transient variation of the fine-structure constant $\a$ as a function of the transient duration, $\t_{\rm int}$.
The secondary horizontal axis shows the corresponding length scale, $d=v_g\t_{\rm int}$.
The shaded curves show the regions of the parameter space that are excluded by various experiments (1$\s$ confidence).
Each curve is valid only below the presented maximum value for $\CT$, the average time between consecutive transients.
The new results of this work are shown in blue.
Existing constraints from optical clock/cavity comparisons are shown in green (Wcis{\l}o {\em et al.}\ \cite{Wcislo2016,Wciso2018}).
Limits also exist from microwave clocks of the GPS constellation (not shown); though they are substantially less stringent ($\delta\a/\a\lesssim10^{-12}$ for $\tau_{\rm int}\sim30\un{s}$) they are valid up to $\CT\simeq16\un{yr}\simeq10^5\un{hr}$~\cite{GPSDM2017}.
}
\label{fig:varalpha}
\end{figure}

\begin{figure*}
\centering
\includegraphics[width=0.99\textwidth]{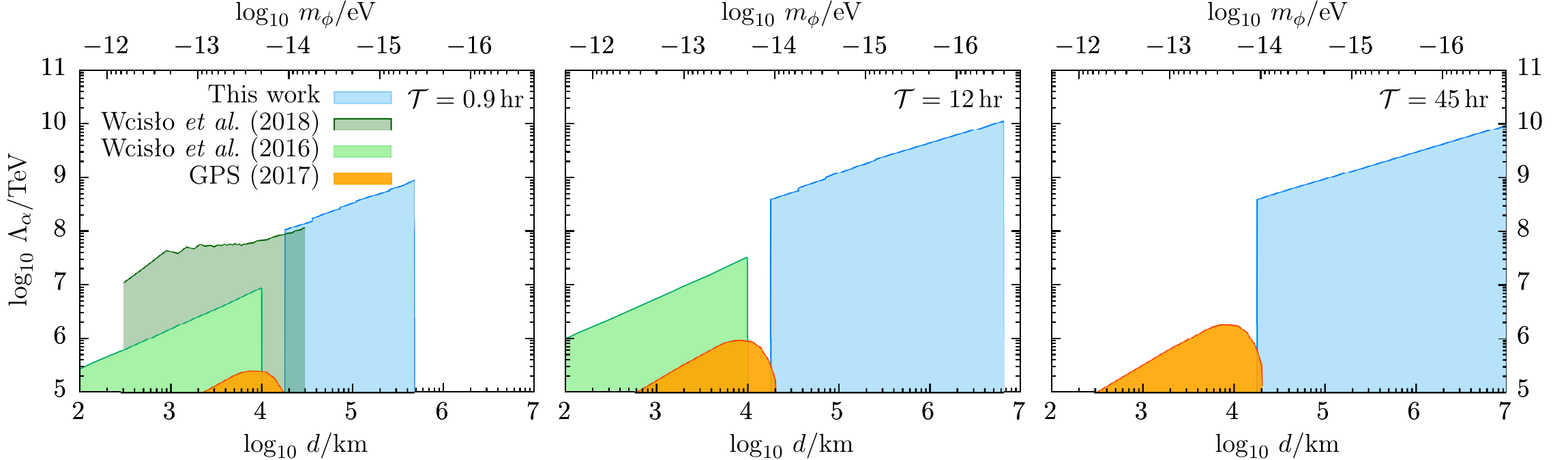}
\caption{Excluded region (1$\s$ confidence) for the effective energy scale $\Lambda_\a$ for topological defect dark matter as a function of the defect size, $d$, for time between events $\CT=0.9\un{hr}$ (left), $\CT=12\un{hr}$ (middle), and $\CT=45\un{hr}$ (right).
The new results from this work are shown in blue, and the existing constraints are shown in green (analysis of optical clock/cavity comparisons in Wcis{\l}o {\em et al.}\ (2016) \cite{Wcislo2016} and (2018) \cite{Wciso2018}) and orange (analysis of the GPS atomic clock data \cite{GPSDM2017}).
These presented $\CT$ values correspond to the maximum applicable for Refs.~\cite{Wciso2018}, \cite{Wcislo2016}, and this work, respectively.
The GPS constraints from Ref.~\cite{GPSDM2017} extend to $\CT\sim10^5\un{hr}$ (they also apply to a combination of interaction parameters, as explained in the text).
}
\label{fig:Lambda}
\end{figure*}


To interpret the analysis in terms of the time between transients, $\CT$, we assume there was (at most) one event during the observation time $T_{\rm obs}$ with magnitude $\delta\a_0^{\rm bf}$, and rule out the possibility of more frequent events with larger magnitudes.
In the analysis, we only use sections of the data that are continuous for periods at least equal to $\t_{\rm int}$ with no gaps.
Therefore, when performing the analysis for larger $\t_{\rm int}$, we are restricted to using less of the data, which reduces the effective observation time.
This reduces the applicable maximum $\CT$ for the largest values of $\t_{\rm int}$ that can be fitted explicitly.
The sensitive region can then be extended beyond this maximum directly probed value to larger $\t_{\rm int}$ according to Eq.~(\ref{eq:dw2}),
so long as $\tau_{\rm int}$ is small compared to both $\CT$ and the total observation time.
These two conditions ensure that the sought signals would be well-separated transients (otherwise they may manifest as roughly constant additions to the clock frequencies, which would not be  observable).
Due to the observation time, we don't extend the constraints beyond $\tau_{\rm int}=10\un{hr}$\,$\simeq\,$4$\times$$10^{4}\un{s}$; this is described in more detail in the Supplemental Material.
For the confidence level, we have assumed that the appearance of the transients follows a Poisson distribution.
We thus place constraints only in the region with average time between transients $\CT<f_P T_{\rm obs}$, where $f_P$ is the Poisson statistics factor ($f_P=0.87$ for a 1$\s$ confidence level).

We first place constraints on transient variations of the fine structure constant, without direct reference to the possible source of the variation.
The results are shown as a function of $\t_{\rm int}$ in Fig.~\ref{fig:varalpha}.
Previous constraints come from optical clock to cavity frequency comparisons \cite{Wcislo2016,Wciso2018}.
There are also complementary constraints from the microwave atomic clocks of the GPS constellation, which apply to a combination of variation in the fine structure constant and the fermion masses \cite{GPSDM2017}.

Our analysis has substantially tightened the constraints on possible transient variations of the fine structure constant, $\a$.
The new constraints are particularly strong for time scales above $\sim$\,$10^2$\,s, where the long-term stability of the atomic clock comparisons in this network offers the largest advantage over existing experiments.
As discussed in the following section, these results also have important implications for the search for dark matter.

\subsection{Topological defect dark matter}

Now, we introduce a specific model that may cause the frequency variations in Eq.~(\ref{eq:dw}).
Consider a scalar field, $\phi$, that has quadratic interactions with standard model particles of the form
\be\label{eq:Lagr}
\CL_{\rm int} = \pm \frac{\phi^2}{\Lambda_\a^2}\frac{1}{4\mu_0}F_{\mu\nu}F^{\mu\nu},
\ee
where $F^{\mu\nu}$ is the electromagnetic Faraday tensor, and $\Lambda_\a$ is the effective energy scale (inverse of the coupling strength).
Such an interaction will lead to the effective rescaling of the fine structure constant, with
\be
\frac{\delta \a(\v{r},t)}{\a} = \pm \frac{\phi^2(\v{r},t)}{\Lambda_\a^2},
\ee
see, e.g., Ref.~\cite{Olive2002}.
If the field $\phi$ has sufficient self interactions, it may form stable macroscopic objects such as topological defects~\cite{Kibble1980,Vilenkin1985}.
The observable variation in $\a$ will occur only when the topological defect overlaps with the clock~\cite{DereviankoDM2014}.

The spatial extent of topological defects is set by the Compton wavelength of the field, $d=\h/(m_\phi c)$, where $m_\phi$ is the field mass.
The energy density inside the defects is $\rho_{\rm inside}=\phi_0^2/(\h cd^2)$, with $\phi_0$ being the maximum value of the field amplitude~\cite{DereviankoDM2014}.
In these models, the field amplitude goes to zero outside the defect.
Assuming that topological defects make up all dark matter, we can link the energy density inside each defect to the average time between events (i.e.\ encounters between a defect and a given point in space):
\be \label{eq:T}
{\CT} 
 =  \frac{\rho_{\rm inside}}{\rho_{\rm DM}}\frac{d}{v_g}
 = \frac{\phi_0^2}{\h c\rho_{\rm DM} v_g d},
\ee
where $\rho_{\rm DM} = (0.3\pm0.1)\un{GeV}\un{cm}^{-3}$~\cite{Bovy:2012tw}
is the galactic dark matter energy density in our solar system~\footnote{For definitiveness, and to be consistent with recent works, we take $\rho_{\rm DM} =0.4\un{GeV}\un{cm}^{-3}$.}.
Combining this with the expression for $\rho_{\rm inside}$
leads to an expression for the field amplitude in terms of the observables and model parameters:
$\phi_0^2 = \h c \, \rho_{\rm DM} v_g \, \CT d$.
We take $d$ and $\CT$
as the free parameters of the model, since they are the direct observables
($\phi_0$, $m_\phi$, and $\rho_{\rm inside}$ are uniquely determined by $d$, $\CT$, and $\rho_{\rm DM}$).

Thereby, the constraints on $\delta\a$ (Fig.~\ref{fig:varalpha}) lead directly to constraints on the effective energy scale $\Lambda_\a$:
\be
\Lambda^2_\a(\CT,d) >
\frac{\h c \rho_{\rm DM} v_g \CT d}{|\delta \a_0(\CT,\t_{\rm int})|/\a}.
\ee
The results are presented in Fig.~\ref{fig:Lambda} as a function of $d$ for a few values of $\CT$.
The constraints reach the $\Lambda_\a$\,$\gtrsim$\,$10^{10}\un{TeV}$ level for $d$\,$\sim$\,$10^7\un{km}$.
Also shown are the existing constraints from atomic clock experiments~\cite{GPSDM2017,Wcislo2016,Wciso2018}.
Other constraints coming from astrophysical observations~\cite{Olive2008,Raffelt1999,Hirata1988} (not shown) are significantly less stringent, and do not exceed the $\sim$\,$10\un{TeV}$ level.

The results from the GPS microwave atomic clocks~\cite{GPSDM2017} (shaded orange in the figures) constrain a combination of interaction parameters, including those stemming from a coupling to fermion masses as well as the coupling to $F_{\mu\nu}^2$ as in Eq.~(\ref{eq:Lagr}).
Therefore, in including those results on the same plot, we are implicitly assuming that the $F_{\mu\nu}$ coupling (leading to effective variation in $\a$) was the dominant coupling for the GPS experiment.

Since we consider long interaction times (i.e.\ large dark matter objects $d\gg L$), all clocks in the network experience essentially the same value of the $\phi$ field.
Therefore  the results presented here apply for topological defects of any geometry (i.e.\ monopoles, strings, or domain walls).
This is in contrast to the results of Refs.~\cite{Wciso2018} and \cite{GPSDM2017}, which explicitly assume a domain wall geometry
(the results of Ref.~\cite{Wcislo2016} also apply for general geometries).
Note also that for such objects to leave transient signatures, they need to be well separated ($\tau_{\rm int}\ll\CT$).
This is equivalent to demanding $\rho_{\rm inside}\gg\rho_{\rm DM}$.
For example, with $d\sim10^4\un{km}$,  Eq.~(\ref{eq:T}) implies that it only makes sense to search for transients with $\CT\gtrsim0.1\un{hr}$.
We also don't extend the limits beyond $\sim$\,$10^7\un{km}$ (corresponding to $\tau_{\rm int}$\,$\sim$\,$10\un{hr}$) for the reasons discussed in the previous section.

\subsection{Conclusion}

By using data from a European network of fiber-linked optical atomic clocks to search for evidence of transient frequency variations, we have substantially improved the constraints on transient variations of the fine structure constant.
With the same analysis we also search for evidence of topological defect dark matter.
At the current sensitivity level, no such evidence was found during the analyzed time windows.
Within the assumptions of our model, we have therefore placed constraints on the possible interactions of such defects with standard model particles, improving upon existing constraints by many orders of magnitude.

We note that it may also be possible to substantially improve the constraints in the region where the event rate is high, $\CT\ll T_{\rm obs}$, even if the signal magnitude is well below the noise, by exploiting statistical signatures~\cite{RobertsAsymm2018}.
For example, in the absence of transients, the distribution of extracted best-fit $\delta\a_0$ values would be expected to be roughly Gaussian.
However, if a large number of transients were present in the data, non-Gaussianities, such as a skewness, would be expected in the distribution.
Further, due to the orbital motion of Earth around the sun in the galactic frame, a $\sim$\,10\% annual modulation in this skewness would also be present if it had a dark matter origin~\cite{RobertsAsymm2018}.
Also, by extending the analysis to lower effective sampling periods, we would have sensitivity to direct measurements of the transient speed and incident direction~\cite{GPSDM2018}, which could be further used to exclude perturbations that cannot be caused by dark matter, and thereby improve the sensitivity of the search in that region.
These avenues will become particularly important as more data and newer experimental techniques become available~\cite{Savalle2019}.

\subsection{Acknowledgements}
BMR gratefully acknowledges financial support from Labex FIRST-TF.
We acknowledge funding support from the Agence Nationale de la Recherche (Labex First-TF ANR-10-LABX-48-01, Equipex REFIMEVE ANR-11-EQPX-0039, Idex PSL ANR-10-IDEX-0001-02).
This work received funding from the European Union's Horizon 2020 program:\ ERC AdOC Grant No.\ 617553.
Funding from the German Research Foundation DFG within the excellence cluster QuantumFrontiers EXC 2123, collaborative research center CRC 1227 (DQ-mat, project B02), collaborative research center CRC 1128 (geo-Q, project A04) and research training group RTG 1729 is acknowledged.
This work has received funding by the European Metrology Programme for Innovation and Research (EMPIR) in project 15SIB03 (OC18) and 15SIB05 (OFTEN). 
The EMPIR initiative is co-funded by the European Union's Horizon 2020 research and innovation programme, and the EMPIR Participating States within EURAMET and the European Union.
NPL authors acknowledge support from the UK Department for Business, Energy and Industrial Strategy as part of the National Measurement System.


\appendix

~\\
\noindent\rule{0.08\textwidth}{0.2pt}\noindent\rule{0.08\textwidth}{1.0pt}\noindent\rule{0.16\textwidth}{1.5pt}%
\noindent\rule{0.08\textwidth}{1.0pt}\noindent\rule{0.08\textwidth}{0.2pt}\\
%
\subsection{Supplemental Material: Data and analysis method}

\begin{figure*}
\includegraphics[width=\textwidth]{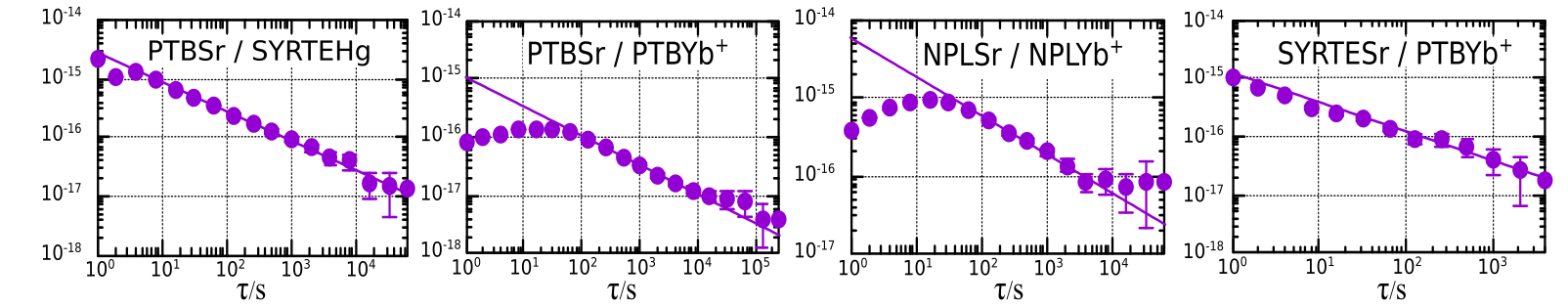}
\caption{\small 
Fractional instabilities for some clock frequency ratios determined by the Allan deviations. 
The solid line shows the $1/\sqrt{\t}$ white-noise trend.
For averaging times $\tau$ larger than $\sim$\,60\,s, the noise is well-modelled as white frequency noise.
}
\label{fig:avar}
\end{figure*}

\begin{figure}
\centering
\includegraphics[width=0.48\textwidth]{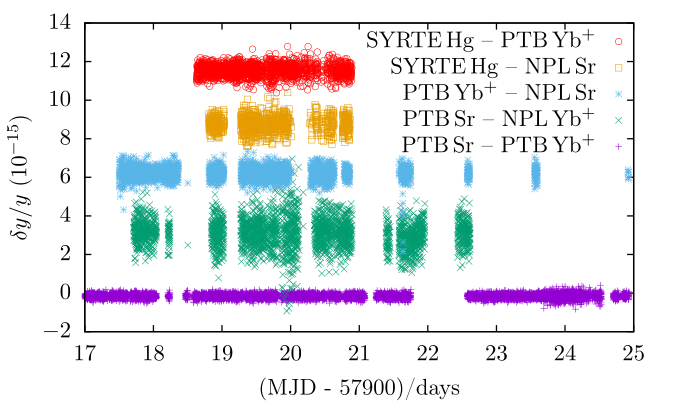}
\caption{\small 
Subset of the clock frequency ratio data (averaged to $60\,$s) from the European fiber-linked optical clock network.
Each time-series is shifted by a constant offset for clarity.
}
\label{fig:data}
\end{figure}

Before the analysis, we average the data into 60\,s bins.
This is done to set the effective sampling period to be greater than the largest servo loop time ($\t=60\,$s), as assumed in Eq.~(\ref{eq:dw}).
We only average over continuous sections of the data, ensuring we do not inadvertently assume any potential signal remains consistent across gaps in the data.
Another effect of this averaging is that above the servo times the data noise can be very well modelled as white frequency noise; see Fig.~\ref{fig:avar}.
An illustrative sub-set of the data is shown in Fig.~\ref{fig:data}.

Also, we restrict our analysis to include only independent clock pairs, so that each clock appears only once in the combined data streams.
The effect of this is to remove any cross correlations between the different clock data streams.
For each separate time window and $\tau_{\rm int}$ value, we choose which clocks to include in order of their effective sensitivity: 
$K_{\rm AB}/\sigma_{y_{AB}}^2$, 
considering only those pairs with continuous data over the given time window.
Here, $\sigma_{y_{AB}}$ is the Allan deviation for the $y_{AB}$ frequency ratio, evaluated at the $60\un{s}$ effective sampling interval.

​We note that​,​ due to the limited frequency width of the atomic transitions, large steps in the cavity--atom frequency difference that last for a sufficiently long period will lead to a loss of lock.
After the source of any such events are identified, the corresponding data are removed.
​​Shorter jumps ​may not lead to the loss of lock, and ​would then​ be indistinguishable from the regular clock noise​. 
Such cases are of not much consequence for our analysis, since we confine our search to longer time periods $\gtrsim\,$60\,s.
In practice, all such events are rare, and their contribution to the downtime of the clocks is negligible. 
At the same time, some very large data outliers are also removed, and are not included in the employed data set.
We note​, however,​ that such large frequency variations cannot be due to the interaction of dark matter with the clock atoms, since such large events would perturb the atomic transition by so much that the laser would lose lock to the atoms, and thus they would not appear in the clock comparisons.

To perform the analysis, we employ a version of the method developed and tested in Ref.~\cite{GPSDM2018}. 
Let $d^i_j$ denote the time series data from the $i$th clock pair at sample-point $j$,
and $\varphi^i_j=\varphi^i_j(\theta)$ denote the ``test signal'' for a given set of model parameters, $\theta$ (e.g.,\ speed, incident direction, coupling strength).
Assuming multi-variate Gaussian likelihoods~\cite{GregoryBayesian2005}, the posterior probability that time window %
$D_{t}$ (centred around time $t$) is consistent with the presence of a (single) transient signal $\varphi$ is
\be\label{eq:pD}
p(D_t|\varphi,\theta) = C \, p(\theta) \exp\left(\tfrac{-1}{2}[d-\varphi]^TH[d-\varphi]\right),
\ee
where $H$ is the inverse of the covariance matrix $E^{ik}_{jl}\equiv\braket{d^i_jd^k_l}$, $p(\theta)$ is the prior probability for the model parameters, and $C$ is a normalisation constant.
In general, the posterior is to be integrated (marginalised) over the model parameters to form the marginal likelihood (evidence).
The signal $\varphi$ can then be calculated according to Eq.~(\ref{eq:dw}) for each of the $N_{\rm cp}$ clock pairs in the network, with the time of arrival of the transient (the time the clock experiences the largest $\delta\a$ magnitude)
determined by the position of each clock and the incident relative velocity of the source of the $\a$-variation.

The posterior for the case that no signal is present in the data (i.e.,\ the data is just noise) is given by Eq.~(\ref{eq:pD}) with $\varphi=0$.
Note that this does not depend on model parameters, so the marginalisation is trivial.
The odds ratio is then given:
\be\label{eq:LogOdds}
O = \int \d\theta \, p(\theta) \,   \exp\left(dH\varphi - \tfrac{1}{2}\varphi H\varphi\right).
\ee
Here, we have used a short-hard notation ($x$ is $d$ or $\varphi$):
\be\label{eq:xHy}
x H \varphi \equiv \sum_{ik}^{N_{\rm cp}}\sum_{jl}^{D_{t}} x^i_j H^{ik}_{jl} \varphi^k_l .
\ee

As noted above, due to the averaging procedure and the inclusion only of independent clock pairs, the data contains essentially no correlations.
In light of this simplification, Eq.~(\ref{eq:xHy}) can be expressed as
\be\label{eq:xHy2}
x H \varphi 
 = \sum_{i}^{N_{\rm cp}} \frac{1}{{\s^i}^2} \sum_{j}^{D_{t}} x^i_j \varphi^i_j,
\ee
where $\s^i$ is the standard deviation of the data from the $i$th clock pair (given by the Allan variance at the 60\,s effective sampling period).

In general, the test signal $\varphi$ depends on the dark matter coupling strengths and the sensitivity of each clock in the network ($K$ factors), as well as the topological defect size, speed, and incident direction.
Then, to calculate the odds ratio, one would integrate over all these parameters taking the Bayesian priors into account, as in Ref.~\cite{GPSDM2018}.
In our case, however, we can make a simplification.
For the considered effective sampling period, $\t=60\,$s, all the clocks in the network can be considered to be co-located (see discussion in the main text).
Therefore, the signal does not depend on the incident direction, and depends only linearly on the speed and coupling strength.
In this case, the odds ratio is maximised simply by maximising the argument of the exponential in Eq.~(\ref{eq:LogOdds}).

We treat the transient duration $\t_{\rm int}$ as a model parameter, and run the analysis separately for each relevant value. 
Noting that the dark matter signal is linear in $\delta\a_0$ (the maximum transient variation in $\a$), we express the test signal as
$
\varphi^i_j \equiv {\delta\a_0} \, s^i_j.
$
Then, the argument of the exponential in Eq.~(\ref{eq:LogOdds}) becomes:
\be\label{eq:CL}
\arg = {\delta\a_0}{}\,dHs - \tfrac{1}{2}\left({\delta\a_0}{}\right)^2 sHs.
\ee
For a given set of parameters, this quantity, and thus the odds ratio (\ref{eq:LogOdds}), is maximised for the best-fit value:
\be\label{eq:hbest}
{\delta\a_0^{\rm bf}} = {dHs}/{sHs}.
\ee
In the absence of a signal, $dHs$ is Gaussian distributed with a mean of zero and a variance equal to $sHs$,
so the ($1\s$) uncertainty in the extracted best-fit is $\Delta\a_0 = (sHs)^{-1/2}$.

\begin{figure}
\centering
\includegraphics[width=0.45\textwidth]{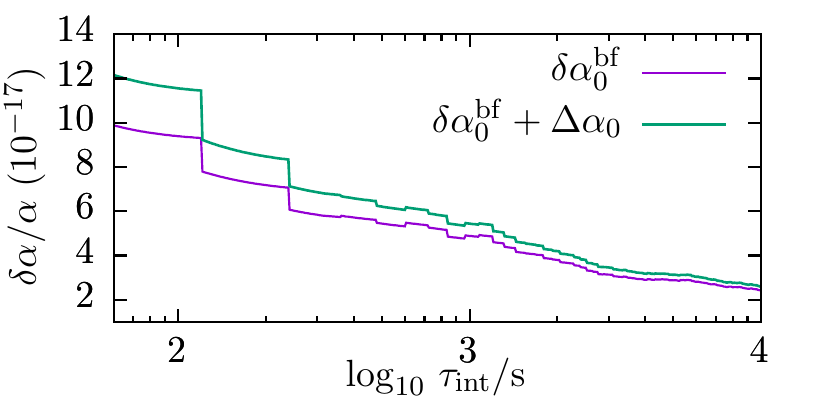}
\caption{The purple line shows the observed maximum best-fit value for $\delta\a_0$ as in Eq.~(\ref{eq:hbest}) as a function of the interaction time, $\t_{\rm int}$. The green line is the $1\s$ confidence bound used to place constraints (\ref{eq:hbest-unc}).
Note that the effective observation time, $T_{\rm obs}$, decreases with increasing $\t_{\rm int}$, since smaller fractions of the data are continuous over the longer time periods, as discussed in the text.}
\label{fig:da-plot}
\end{figure}

The best-fit $\delta\a_0$ (\ref{eq:hbest}) is then calculated as a function of time (and $\t_{\rm int}$) over the span of the data.
By this we mean that we calculate the best-fit over a given time window, and then step this window along by the smallest available increment, $\tau_0$.
The windows are assumed to be centred on the (possible) transient incident time, and extend to cover at least a time period equal to $\tau_{\rm int}$.
We tested several values, and found that exactly how large each window is makes essentially no difference to the results (since the signal template $s$ goes to zero quickly outside this region).
For each $\t_{\rm int}$, the largest best-fit value  found throughout the entire observation time can be used to place constraints: 
\be\label{eq:hbest-unc}
| {\delta\a_0}| < |{\delta\a_0^{\rm bf}}|_{\rm max} + n_{\rm CL}(sHs)^{-1/2},
\ee
where $n_{\rm CL}=1$ for 1$\s$ confidence.
We consider only time periods when at least two clock pairs (four clocks) were actively taking data. 
For a given interaction duration, we only include data streams which have continuous data (i.e.,\ sampled every $1\,$s up to at least the considered $\t_{\rm int}$).
This means that the effective observation time, $T_{\rm obs}$, decreases with increasing $\t_{\rm int}$.
For $10^2\un{s}$, we have $T_{\rm obs}=47\un{hr}$, while for $10^3\un{s}$ we have $T_{\rm obs}=15\un{hr}$.
The best fit values, and the 1$\s$ confidence bound, are shown in Fig.~\ref{fig:da-plot}.

By finding the largest $\delta\a_0$ value that appears in the data, we are assuming there was (at most) one event present in the data with magnitude $\delta\a_0$, and then ruling out the possibility of events with magnitudes larger than this (at the stated confidence level).
These constraints then apply to the parameter space region for time between events $\CT<f_PT_{\rm obs}$ (where $f_P<1$ is the Poisson statistics factor).
This is the most conservative approach.
It may be possible to set more stringent limits applicable for lower $\CT$ values by finding the largest values for $\delta\a_0$ that appear in the data at least $n=f_P^{(n)} T_{\rm obs}/\CT$ times.

We search through each $\t_{\rm int}$ specifically between a minimum and maximum value, which are set respectively by the effective sampling period (60\,s) and the longest stretch of continuous data in the current data set ($\tau_{\rm max}\sim10^4\,$s).
Assuming there is no true $\delta\a$ signal in the data, the observed maximum frequency variations that last for duration $\t_{\rm int}$ would be expected to scale as $\delta{y}/{y}\propto\sigma/\sqrt{\tau_{\rm int}}$.
Therefore, between the maximum and minimum directly tested values, the constraints are expected to scale as $\sqrt{\t_{\rm int}}$, which is seen in the results.

A transient with $\t_{\rm int}\gg\t_{\rm max}$ will leave a signal in the data that is approximately constant over the $\t_{\rm max}$ period (\ref{eq:dw2}).
Therefore, by performing a fit to $\delta\a_0$ in this case, we can search for evidence of transients with very large $\tau_{\rm int}$.
However, in this case, the sensitivity does not increase with increasing $\t_{\rm int}$ as in the $<\t_{\rm max}$ case, but instead stays constant, see Eq.~(\ref{eq:dw2}).
We note also, that this procedure does not extend the sensitivity indefinitely as $\t_{\rm int}\to\infty$.
In order to measure a transient frequency variation, one must know the ``real'', or long-term average, frequency from which it varies.
It only makes sense to claim knowledge of the unperturbed ratio $y_{AB}$ if the total measurement time is much greater than $\t_{\rm int}$.
We therefore do not extend the constraints past $\t_{\rm int}\sim4\times
10^4\un{s}\sim10\un{hr}$, which is about $50\%$ of the total for the clock pair with the shortest measurement duration, and about $5\%$ of that for the longest.
In reality, the constraints are typically bounded well before this value due to the condition that the transients be well-separated, i.e.,\ $\tau_{\rm int}\ll\mathcal{T}$ (which we take as $\tau_{\rm int} < \mathcal{T}/5$).

\begin{figure}
\centering
\includegraphics[width=0.45\textwidth]{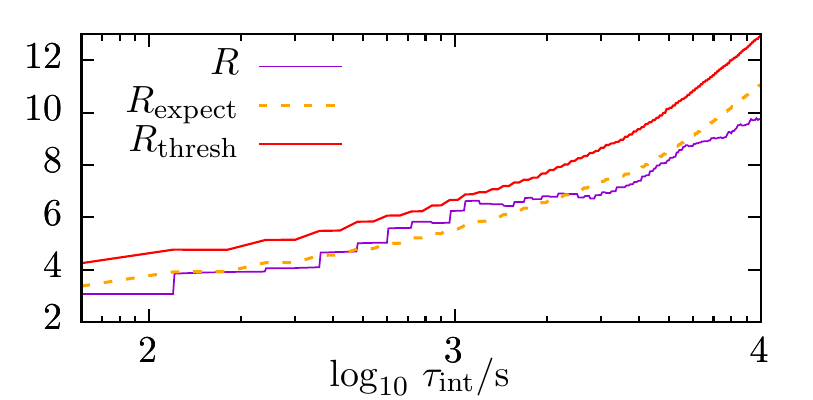}
\caption{The purple line shows the maximum value for $R$ calculated from the current data set as in Eq.~(\ref{eq:R}) as a function of the interaction time, $\t_{\rm int}$. The red line is the threshold, above which statistical false-positives do not occur at 99\% confidence, and the dashed orange line is the expected value for $R$ in the absence of a signal. Both $R_{\rm thresh}$ and $R_{\rm expect}$ are calculated from simulations mimicking the current data set, assuming white frequency noise.}
\label{fig:RThresh}
\end{figure}

To search for potential positive events, instead of maximising the best-fit value for $\delta\a_0$, we maximise the likelihood itself.
In our case, this is equivalent to maximising the argument in Eq.~(\ref{eq:CL}).
Substituting $\delta\a_0^{\rm bf}$ (\ref{eq:hbest}) back into (\ref{eq:CL}) gives the value that maximises the likelihood.
For convenience, we take the square root of this quantity, and define
\be\label{eq:R}
R \equiv \frac{dHs}{\sqrt{2sHs}}.
\ee
Note that $R$ has the form of a signal to noise ratio. In fact  it is equivalent to the ratio $\delta\a_0^{\rm bf}/\Delta\a_0$ (up to a constant factor).
As before, for a given $\tau_{\rm int}$, we find that maximum value of $R$ that occurs throughout the observation time.

To determine the significance of any potentially detected event, we define a threshold, $R_{\rm thresh}$, above which it is sufficiently unlikely that there is a false positive due to random noise processes.
To determine the threshold, we use a Monte-Carlo procedure, generating random white noise according to the known average noise levels for each clock pair.
The data is generated in such a way as to match the characteristics of the networks, i.e.,\ which clock pairs were running at which times, including emulating any gaps in the data time-series.
This simulated data is then run through the exact same search method described above, and we record the maximum $R$ values as a function of $\t_{\rm int}$.
We repeat this process a large number of times (1000), and determine the level at which there are no statistical false positives at the 99\% confidence level
(false positive is defined here as any time $|R|>R_{\rm thresh}$ due to purely random noise processes).
At the same time, we also define the expected value, $R_{\rm expect}$, which is calculated as the mean of the maximum $R$ value for each $\tau_{\rm int}$ extracted from the same simulations assuming white frequency noise.

The maximum extracted $R$ values are shown as a function of $\t_{\rm int}$ in Fig.~\ref{fig:RThresh}, along with the calculated threshold, $R_{\rm thresh}$.
There are regions (around $\tau_{\rm int}\sim10^3\,$s) where the observed $R$ value exceeds that expected for white frequency noise in the absence of a signal, however the significance is low.
Using the considered data set, we find no occurrences where the likelihood exceeds the threshold at the current sensitivity level.

\bibliography{transient}

\begin{thebibliography}{48}%
\makeatletter
\providecommand \@ifxundefined [1]{%
 \@ifx{#1\undefined}
}%
\providecommand \@ifnum [1]{%
 \ifnum #1\expandafter \@firstoftwo
 \else \expandafter \@secondoftwo
 \fi
}%
\providecommand \@ifx [1]{%
 \ifx #1\expandafter \@firstoftwo
 \else \expandafter \@secondoftwo
 \fi
}%
\providecommand \natexlab [1]{#1}%
\providecommand \enquote  [1]{``#1''}%
\providecommand \bibnamefont  [1]{#1}%
\providecommand \bibfnamefont [1]{#1}%
\providecommand \citenamefont [1]{#1}%
\providecommand \href@noop [0]{\@secondoftwo}%
\providecommand \href [0]{\begingroup \@sanitize@url \@href}%
\providecommand \@href[1]{\@@startlink{#1}\@@href}%
\providecommand \@@href[1]{\endgroup#1\@@endlink}%
\providecommand \@sanitize@url [0]{\catcode `\\12\catcode `\$12\catcode
  `\&12\catcode `\#12\catcode `\^12\catcode `\_12\catcode `\%12\relax}%
\providecommand \@@startlink[1]{}%
\providecommand \@@endlink[0]{}%
\providecommand \url  [0]{\begingroup\@sanitize@url \@url }%
\providecommand \@url [1]{\endgroup\@href {#1}{\urlprefix }}%
\providecommand \urlprefix  [0]{URL }%
\providecommand \Eprint [0]{\href }%
\providecommand \doibase [0]{http://dx.doi.org/}%
\providecommand \selectlanguage [0]{\@gobble}%
\providecommand \bibinfo  [0]{\@secondoftwo}%
\providecommand \bibfield  [0]{\@secondoftwo}%
\providecommand \translation [1]{[#1]}%
\providecommand \BibitemOpen [0]{}%
\providecommand \bibitemStop [0]{}%
\providecommand \bibitemNoStop [0]{.\EOS\space}%
\providecommand \EOS [0]{\spacefactor3000\relax}%
\providecommand \BibitemShut  [1]{\csname bibitem#1\endcsname}%
\let\auto@bib@innerbib\@empty
\bibitem [{\citenamefont {Liu}\ \emph {et~al.}(2017)\citenamefont {Liu},
  \citenamefont {Chen}, and \citenamefont {Ji}}]{Liu2017}%
  \BibitemOpen
  \bibfield  {author} {\bibinfo {author} {\bibfnamefont {J.}~\bibnamefont
  {Liu}}, \bibinfo {author} {\bibfnamefont {X.}~\bibnamefont {Chen}},  and
  \bibinfo {author} {\bibfnamefont {X.}~\bibnamefont {Ji}},  }\href {\doibase
  10.1038/nphys4039} {\bibfield  {journal} {\bibinfo  {journal} {Nat. Phys.}
  }\textbf {\bibinfo {volume} {13}},  \bibinfo {pages} {212} (\bibinfo {year}
  {2017})}\BibitemShut {NoStop}%
\bibitem [{\citenamefont {Bertone} and \citenamefont
  {Tait}(2018)}]{Bertone2018}%
  \BibitemOpen
  \bibfield  {author} {\bibinfo {author} {\bibfnamefont {G.}~\bibnamefont
  {Bertone}} and \bibinfo {author} {\bibfnamefont {T.~M.~P.}\ \bibnamefont
  {Tait}},  }\href {\doibase 10.1038/s41586-018-0542-z} {\bibfield  {journal}
  {\bibinfo  {journal} {Nature}\ }\textbf {\bibinfo {volume} {562}},  \bibinfo
  {pages} {51} (\bibinfo {year} {2018})},  \Eprint
  {http://arxiv.org/abs/1810.01668} {arXiv:1810.01668} \BibitemShut {NoStop}%
\bibitem [{\citenamefont {Preskill}\ \emph {et~al.}(1983)\citenamefont
  {Preskill}, \citenamefont {Wise}, and \citenamefont
  {Wilczek}}]{Preskill1983}%
  \BibitemOpen
  \bibfield  {author} {\bibinfo {author} {\bibfnamefont {J.}~\bibnamefont
  {Preskill}}, \bibinfo {author} {\bibfnamefont {M.~B.}\ \bibnamefont {Wise}},
   and \bibinfo {author} {\bibfnamefont {F.}~\bibnamefont {Wilczek}},  }\href
  {\doibase 10.1016/0370-2693(83)90637-8} {\bibfield  {journal} {\bibinfo
  {journal} {Phys. Lett. B}\ }\textbf {\bibinfo {volume} {120}},  \bibinfo
  {pages} {127} (\bibinfo {year} {1983})}\BibitemShut {NoStop}%
\bibitem [{\citenamefont {Dine} and \citenamefont
  {Fischler}(1983)}]{Dine1983}%
  \BibitemOpen
  \bibfield  {author} {\bibinfo {author} {\bibfnamefont {M.}~\bibnamefont
  {Dine}} and \bibinfo {author} {\bibfnamefont {W.}~\bibnamefont
  {Fischler}},  }\href {\doibase 10.1016/0370-2693(83)90639-1} {\bibfield
  {journal} {\bibinfo  {journal} {Phys. Lett. B}\ }\textbf {\bibinfo {volume}
  {120}},  \bibinfo {pages} {137} (\bibinfo {year} {1983})}\BibitemShut
  {NoStop}%
\bibitem [{\citenamefont {Kibble}(1980)}]{Kibble1980}%
  \BibitemOpen
  \bibfield  {author} {\bibinfo {author} {\bibfnamefont {T.~W.~B.}
  \bibnamefont {Kibble}},  }\href {\doibase 10.1016/0370-1573(80)90091-5}
  {\bibfield  {journal} {\bibinfo  {journal} {Phys. Rep.}\ }\textbf {\bibinfo
  {volume} {67}},  \bibinfo {pages} {183} (\bibinfo {year} {1980})}\BibitemShut
  {NoStop}%
\bibitem [{\citenamefont {Vilenkin}(1985)}]{Vilenkin1985}%
  \BibitemOpen
  \bibfield  {author} {\bibinfo {author} {\bibfnamefont {A.}~\bibnamefont
  {Vilenkin}},  }\href {\doibase 10.1016/0370-1573(85)90033-X} {\bibfield
  {journal} {\bibinfo  {journal} {Phys. Rep.}\ }\textbf {\bibinfo {volume}
  {121}},  \bibinfo {pages} {263} (\bibinfo {year} {1985})}\BibitemShut
  {NoStop}%
\bibitem [{\citenamefont {Derevianko} and \citenamefont
  {Pospelov}(2014)}]{DereviankoDM2014}%
  \BibitemOpen
  \bibfield  {author} {\bibinfo {author} {\bibfnamefont {A.}~\bibnamefont
  {Derevianko}} and \bibinfo {author} {\bibfnamefont {M.}~\bibnamefont
  {Pospelov}},  }\href {\doibase 10.1038/nphys3137} {\bibfield  {journal}
  {\bibinfo  {journal} {Nat. Phys.}\ }\textbf {\bibinfo {volume} {10}},
  \bibinfo {pages} {933} (\bibinfo {year} {2014})}\BibitemShut {NoStop}%
\bibitem [{\citenamefont {Safronova}\ \emph {et~al.}(2018)\citenamefont
  {Safronova}, \citenamefont {Budker}, \citenamefont {DeMille}, \citenamefont
  {Kimball}, \citenamefont {Derevianko}, and \citenamefont
  {Clark}}]{AtomicReview2017}%
  \BibitemOpen
  \bibfield  {author} {\bibinfo {author} {\bibfnamefont {M.~S.}\ \bibnamefont
  {Safronova}}, \bibinfo {author} {\bibfnamefont {D.}~\bibnamefont {Budker}},
  \bibinfo {author} {\bibfnamefont {D.}~\bibnamefont {DeMille}}, \bibinfo
  {author} {\bibfnamefont {D.~F.~J.}\ \bibnamefont {Kimball}}, \bibinfo
  {author} {\bibfnamefont {A.}~\bibnamefont {Derevianko}},  and \bibinfo
  {author} {\bibfnamefont {C.~W.}\ \bibnamefont {Clark}},  }\href {\doibase
  10.1103/RevModPhys.90.025008} {\bibfield  {journal} {\bibinfo  {journal}
  {Rev. Mod. Phys.}\ }\textbf {\bibinfo {volume} {90}},  \bibinfo {pages}
  {025008} (\bibinfo {year} {2018})}\BibitemShut {NoStop}%
\bibitem [{\citenamefont {Olive} and \citenamefont
  {Pospelov}(2002)}]{Olive2002}%
  \BibitemOpen
  \bibfield  {author} {\bibinfo {author} {\bibfnamefont {K.~A.}\ \bibnamefont
  {Olive}} and \bibinfo {author} {\bibfnamefont {M.}~\bibnamefont
  {Pospelov}},  }\href {\doibase 10.1103/PhysRevD.65.085044} {\bibfield
  {journal} {\bibinfo  {journal} {Phys. Rev. D}\ }\textbf {\bibinfo {volume}
  {65}},  \bibinfo {pages} {085044} (\bibinfo {year} {2002})}\BibitemShut
  {NoStop}%
\bibitem [{\citenamefont {Stadnik} and \citenamefont
  {Flambaum}(2015)}]{StadnikDMalpha2015}%
  \BibitemOpen
  \bibfield  {author} {\bibinfo {author} {\bibfnamefont {Y.~V.}\ \bibnamefont
  {Stadnik}} and \bibinfo {author} {\bibfnamefont {V.~V.}\ \bibnamefont
  {Flambaum}},  }\href {\doibase 10.1103/PhysRevLett.115.201301} {\bibfield
  {journal} {\bibinfo  {journal} {Phys. Rev. Lett.}\ }\textbf {\bibinfo
  {volume} {115}},  \bibinfo {pages} {201301} (\bibinfo {year}
  {2015})}\BibitemShut {NoStop}%
\bibitem [{\citenamefont {Godun}\ \emph {et~al.}(2014)\citenamefont {Godun},
  \citenamefont {Nisbet-Jones}, \citenamefont {Jones}, \citenamefont {King},
  \citenamefont {Johnson}, \citenamefont {Margolis}, \citenamefont {Szymaniec},
  \citenamefont {Lea}, \citenamefont {Bongs}, and \citenamefont
  {Gill}}]{GodNisJon14}%
  \BibitemOpen
  \bibfield  {author} {\bibinfo {author} {\bibfnamefont {R.~M.}\ \bibnamefont
  {Godun}}, \bibinfo {author} {\bibfnamefont {P.~B.~R.}\ \bibnamefont
  {Nisbet-Jones}}, \bibinfo {author} {\bibfnamefont {J.~M.}\ \bibnamefont
  {Jones}}, \bibinfo {author} {\bibfnamefont {S.~A.}\ \bibnamefont {King}},
  \bibinfo {author} {\bibfnamefont {L.~A.~M.}\ \bibnamefont {Johnson}},
  \bibinfo {author} {\bibfnamefont {H.~S.}\ \bibnamefont {Margolis}}, \bibinfo
  {author} {\bibfnamefont {K.}~\bibnamefont {Szymaniec}}, \bibinfo {author}
  {\bibfnamefont {S.~N.}\ \bibnamefont {Lea}}, \bibinfo {author} {\bibfnamefont
  {K.}~\bibnamefont {Bongs}},  and \bibinfo {author} {\bibfnamefont
  {P.}~\bibnamefont {Gill}},  }\href {\doibase 10.1103/PhysRevLett.113.210801}
  {\bibfield  {journal} {\bibinfo  {journal} {Phys. Rev. Lett.}\ }\textbf
  {\bibinfo {volume} {113}},  \bibinfo {pages} {210801} (\bibinfo {year}
  {2014})}\BibitemShut {NoStop}%
\bibitem [{\citenamefont {Huntemann}\ \emph {et~al.}(2014)\citenamefont
  {Huntemann}, \citenamefont {Lipphardt}, \citenamefont {Tamm}, \citenamefont
  {Gerginov}, \citenamefont {Weyers}, and \citenamefont
  {Peik}}]{HunLipTam14}%
  \BibitemOpen
  \bibfield  {author} {\bibinfo {author} {\bibfnamefont {N.}~\bibnamefont
  {Huntemann}}, \bibinfo {author} {\bibfnamefont {B.}~\bibnamefont
  {Lipphardt}}, \bibinfo {author} {\bibfnamefont {C.}~\bibnamefont {Tamm}},
  \bibinfo {author} {\bibfnamefont {V.}~\bibnamefont {Gerginov}}, \bibinfo
  {author} {\bibfnamefont {S.}~\bibnamefont {Weyers}},  and \bibinfo {author}
  {\bibfnamefont {E.}~\bibnamefont {Peik}},  }\href {\doibase
  10.1103/PhysRevLett.113.210802} {\bibfield  {journal} {\bibinfo  {journal}
  {Phys. Rev. Lett.}\ }\textbf {\bibinfo {volume} {113}},  \bibinfo {pages}
  {210802} (\bibinfo {year} {2014})}\BibitemShut {NoStop}%
\bibitem [{\citenamefont {McGrew}\ \emph {et~al.}(2019)\citenamefont {McGrew},
  \citenamefont {Zhang}, \citenamefont {Leopardi}, \citenamefont {Fasano},
  \citenamefont {Nicolodi}, \citenamefont {Beloy}, \citenamefont {Yao},
  \citenamefont {Sherman}, \citenamefont {Sch{\"{a}}ffer}, \citenamefont
  {Savory}, \citenamefont {Brown}, \citenamefont {R{\"{o}}misch}, \citenamefont
  {Oates}, \citenamefont {Parker}, \citenamefont {Fortier}, and \citenamefont
  {Ludlow}}]{McGrew2018a}%
  \BibitemOpen
  \bibfield  {author} {\bibinfo {author} {\bibfnamefont {W.~F.}\ \bibnamefont
  {McGrew}}, \bibinfo {author} {\bibfnamefont {X.}~\bibnamefont {Zhang}},
  \bibinfo {author} {\bibfnamefont {H.}~\bibnamefont {Leopardi}}, \bibinfo
  {author} {\bibfnamefont {R.~J.}\ \bibnamefont {Fasano}}, \bibinfo {author}
  {\bibfnamefont {D.}~\bibnamefont {Nicolodi}}, \bibinfo {author}
  {\bibfnamefont {K.}~\bibnamefont {Beloy}}, \bibinfo {author} {\bibfnamefont
  {J.}~\bibnamefont {Yao}}, \bibinfo {author} {\bibfnamefont {J.~A.}
  \bibnamefont {Sherman}}, \bibinfo {author} {\bibfnamefont {S.~A.}
  \bibnamefont {Sch{\"{a}}ffer}}, \bibinfo {author} {\bibfnamefont
  {J.}~\bibnamefont {Savory}}, \bibinfo {author} {\bibfnamefont {R.~C.}
  \bibnamefont {Brown}}, \bibinfo {author} {\bibfnamefont {S.}~\bibnamefont
  {R{\"{o}}misch}}, \bibinfo {author} {\bibfnamefont {C.~W.}\ \bibnamefont
  {Oates}}, \bibinfo {author} {\bibfnamefont {T.~E.}\ \bibnamefont {Parker}},
  \bibinfo {author} {\bibfnamefont {T.~M.}\ \bibnamefont {Fortier}},  and
  \bibinfo {author} {\bibfnamefont {A.~D.}\ \bibnamefont {Ludlow}},  }\href
  {\doibase 10.1364/OPTICA.6.000448} {\bibfield  {journal} {\bibinfo  {journal}
  {Optica}\ }\textbf {\bibinfo {volume} {6}},  \bibinfo {pages} {448} (\bibinfo
  {year} {2019})},  \Eprint {http://arxiv.org/abs/1811.05885}
  {arXiv:1811.05885} \BibitemShut {NoStop}%
\bibitem [{\citenamefont {Arvanitaki}\ \emph {et~al.}(2015)\citenamefont
  {Arvanitaki}, \citenamefont {Huang}, and \citenamefont {{Van
  Tilburg}}}]{Arvanitaki2014}%
  \BibitemOpen
  \bibfield  {author} {\bibinfo {author} {\bibfnamefont {A.}~\bibnamefont
  {Arvanitaki}}, \bibinfo {author} {\bibfnamefont {J.}~\bibnamefont {Huang}}, 
  and \bibinfo {author} {\bibfnamefont {K.}~\bibnamefont {{Van Tilburg}}},
  }\href {\doibase 10.1103/PhysRevD.91.015015} {\bibfield  {journal} {\bibinfo
  {journal} {Phys. Rev. D}\ }\textbf {\bibinfo {volume} {91}},  \bibinfo
  {pages} {015015} (\bibinfo {year} {2015})}\BibitemShut {NoStop}%
\bibitem [{\citenamefont {{Van Tilburg}}\ \emph {et~al.}(2015)\citenamefont
  {{Van Tilburg}}, \citenamefont {Leefer}, \citenamefont {Bougas}, and
  \citenamefont {Budker}}]{Tilburg2015}%
  \BibitemOpen
  \bibfield  {author} {\bibinfo {author} {\bibfnamefont {K.}~\bibnamefont {{Van
  Tilburg}}}, \bibinfo {author} {\bibfnamefont {N.}~\bibnamefont {Leefer}},
  \bibinfo {author} {\bibfnamefont {L.}~\bibnamefont {Bougas}},  and \bibinfo
  {author} {\bibfnamefont {D.}~\bibnamefont {Budker}},  }\href {\doibase
  10.1103/PhysRevLett.115.011802} {\bibfield  {journal} {\bibinfo  {journal}
  {Phys. Rev. Lett.}\ }\textbf {\bibinfo {volume} {115}},  \bibinfo {pages}
  {011802} (\bibinfo {year} {2015})}\BibitemShut {NoStop}%
\bibitem [{\citenamefont {Hees}\ \emph {et~al.}(2016)\citenamefont {Hees},
  \citenamefont {Gu{\'{e}}na}, \citenamefont {Abgrall}, \citenamefont {Bize},
  and \citenamefont {Wolf}}]{Hees2016}%
  \BibitemOpen
  \bibfield  {author} {\bibinfo {author} {\bibfnamefont {A.}~\bibnamefont
  {Hees}}, \bibinfo {author} {\bibfnamefont {J.}~\bibnamefont {Gu{\'{e}}na}},
  \bibinfo {author} {\bibfnamefont {M.}~\bibnamefont {Abgrall}}, \bibinfo
  {author} {\bibfnamefont {S.}~\bibnamefont {Bize}},  and \bibinfo {author}
  {\bibfnamefont {P.}~\bibnamefont {Wolf}},  }\href {\doibase
  10.1103/PhysRevLett.117.061301} {\bibfield  {journal} {\bibinfo  {journal}
  {Phys. Rev. Lett.}\ }\textbf {\bibinfo {volume} {117}},  \bibinfo {pages}
  {061301} (\bibinfo {year} {2016})}\BibitemShut {NoStop}%
\bibitem [{\citenamefont {Hees}\ \emph {et~al.}(2018)\citenamefont {Hees},
  \citenamefont {Minazzoli}, \citenamefont {Savalle}, \citenamefont {Stadnik},
  and \citenamefont {Wolf}}]{Hees2018}%
  \BibitemOpen
  \bibfield  {author} {\bibinfo {author} {\bibfnamefont {A.}~\bibnamefont
  {Hees}}, \bibinfo {author} {\bibfnamefont {O.}~\bibnamefont {Minazzoli}},
  \bibinfo {author} {\bibfnamefont {E.}~\bibnamefont {Savalle}}, \bibinfo
  {author} {\bibfnamefont {Y.~V.}\ \bibnamefont {Stadnik}},  and \bibinfo
  {author} {\bibfnamefont {P.}~\bibnamefont {Wolf}},  }\href {\doibase
  10.1103/PhysRevD.98.064051} {\bibfield  {journal} {\bibinfo  {journal} {Phys.
  Rev. D}\ }\textbf {\bibinfo {volume} {98}},  \bibinfo {pages} {064051}
  (\bibinfo {year} {2018})}\BibitemShut {NoStop}%
\bibitem [{\citenamefont {Derevianko}(2018)}]{DereviankoVULF2016}%
  \BibitemOpen
  \bibfield  {author} {\bibinfo {author} {\bibfnamefont {A.}~\bibnamefont
  {Derevianko}},  }\href {\doibase 10.1103/PhysRevA.97.042506} {\bibfield
  {journal} {\bibinfo  {journal} {Phys. Rev. A}\ }\textbf {\bibinfo {volume}
  {97}},  \bibinfo {pages} {042506} (\bibinfo {year} {2018})}\BibitemShut
  {NoStop}%
\bibitem [{\citenamefont {Geraci}\ \emph {et~al.}(2018)\citenamefont {Geraci},
  \citenamefont {Bradley}, \citenamefont {Gao}, \citenamefont {Weinstein},
  and \citenamefont {Derevianko}}]{Geraci2018}%
  \BibitemOpen
  \bibfield  {author} {\bibinfo {author} {\bibfnamefont {A.~A.}\ \bibnamefont
  {Geraci}}, \bibinfo {author} {\bibfnamefont {C.}~\bibnamefont {Bradley}},
  \bibinfo {author} {\bibfnamefont {D.}~\bibnamefont {Gao}}, \bibinfo {author}
  {\bibfnamefont {J.}~\bibnamefont {Weinstein}},  and \bibinfo {author}
  {\bibfnamefont {A.}~\bibnamefont {Derevianko}},  }\href{http://arxiv.org/abs/1808.00540} {   (\bibinfo {year} {2018})},  \Eprint
  {http://arxiv.org/abs/1808.00540} {arXiv:1808.00540} \BibitemShut {NoStop}%
\bibitem [{\citenamefont {Wolf}\ \emph {et~al.}(2019)\citenamefont {Wolf},
  \citenamefont {Alonso}, and \citenamefont {Blas}}]{Wolf2018a}%
  \BibitemOpen
  \bibfield  {author} {\bibinfo {author} {\bibfnamefont {P.}~\bibnamefont
  {Wolf}}, \bibinfo {author} {\bibfnamefont {R.}~\bibnamefont {Alonso}},  and
  \bibinfo {author} {\bibfnamefont {D.}~\bibnamefont {Blas}},  }\href {\doibase
  10.1103/PhysRevD.99.095019} {\bibfield  {journal} {\bibinfo  {journal} {Phys.
  Rev. D}\ }\textbf {\bibinfo {volume} {99}},  \bibinfo {pages} {095019}
  (\bibinfo {year} {2019})},  \Eprint {http://arxiv.org/abs/1810.01632}
  {arXiv:1810.01632} \BibitemShut {NoStop}%
\bibitem [{\citenamefont {Pospelov}\ \emph {et~al.}(2013)\citenamefont
  {Pospelov}, \citenamefont {Pustelny}, \citenamefont {Ledbetter},
  \citenamefont {Kimball}, \citenamefont {Gawlik}, and \citenamefont
  {Budker}}]{Pospelov2013}%
  \BibitemOpen
  \bibfield  {author} {\bibinfo {author} {\bibfnamefont {M.}~\bibnamefont
  {Pospelov}}, \bibinfo {author} {\bibfnamefont {S.}~\bibnamefont {Pustelny}},
  \bibinfo {author} {\bibfnamefont {M.~P.}\ \bibnamefont {Ledbetter}}, \bibinfo
  {author} {\bibfnamefont {D.~F.~J.}\ \bibnamefont {Kimball}}, \bibinfo
  {author} {\bibfnamefont {W.}~\bibnamefont {Gawlik}},  and \bibinfo {author}
  {\bibfnamefont {D.}~\bibnamefont {Budker}},  }\href {\doibase
  10.1103/PhysRevLett.110.021803} {\bibfield  {journal} {\bibinfo  {journal}
  {Phys. Rev. Lett.}\ }\textbf {\bibinfo {volume} {110}},  \bibinfo {pages}
  {021803} (\bibinfo {year} {2013})}\BibitemShut {NoStop}%
\bibitem [{\citenamefont {Alonso}\ \emph {et~al.}(2018)\citenamefont {Alonso},
  \citenamefont {Blas}, and \citenamefont {Wolf}}]{Alonso2018}%
  \BibitemOpen
  \bibfield  {author} {\bibinfo {author} {\bibfnamefont {R.}~\bibnamefont
  {Alonso}}, \bibinfo {author} {\bibfnamefont {D.}~\bibnamefont {Blas}},  and
  \bibinfo {author} {\bibfnamefont {P.}~\bibnamefont {Wolf}},  }\href
  {http://arxiv.org/abs/1810.00889} {   (\bibinfo {year} {2018})},  \Eprint
  {http://arxiv.org/abs/1810.00889} {arXiv:1810.00889} \BibitemShut {NoStop}%
\bibitem [{\citenamefont {Olive} and \citenamefont
  {Pospelov}(2008)}]{Olive2008}%
  \BibitemOpen
  \bibfield  {author} {\bibinfo {author} {\bibfnamefont {K.~A.}\ \bibnamefont
  {Olive}} and \bibinfo {author} {\bibfnamefont {M.}~\bibnamefont
  {Pospelov}},  }\href {\doibase 10.1103/PhysRevD.77.043524} {\bibfield
  {journal} {\bibinfo  {journal} {Phys. Rev. D}\ }\textbf {\bibinfo {volume}
  {77}},  \bibinfo {pages} {043524} (\bibinfo {year} {2008})}\BibitemShut
  {NoStop}%
\bibitem [{\citenamefont {Flambaum} and \citenamefont
  {Dzuba}(2009)}]{FlambaumCJP2009}%
  \BibitemOpen
  \bibfield  {author} {\bibinfo {author} {\bibfnamefont {V.~V.}\ \bibnamefont
  {Flambaum}} and \bibinfo {author} {\bibfnamefont {V.~A.}\ \bibnamefont
  {Dzuba}},  }\href {\doibase 10.1139/p08-072} {\bibfield  {journal} {\bibinfo
  {journal} {Can. J. Phys.}\ }\textbf {\bibinfo {volume} {87}},  \bibinfo
  {pages} {25} (\bibinfo {year} {2009})}\BibitemShut {NoStop}%
\bibitem [{\citenamefont {Dzuba}\ \emph
  {et~al.}(1999{\natexlab{a}})\citenamefont {Dzuba}, \citenamefont {Flambaum},
  and \citenamefont {Webb}}]{Dzuba1999}%
  \BibitemOpen
  \bibfield  {author} {\bibinfo {author} {\bibfnamefont {V.~A.}\ \bibnamefont
  {Dzuba}}, \bibinfo {author} {\bibfnamefont {V.~V.}\ \bibnamefont {Flambaum}},
   and \bibinfo {author} {\bibfnamefont {J.~K.}\ \bibnamefont {Webb}},
  }\href {\doibase 10.1103/PhysRevA.59.230} {\bibfield  {journal} {\bibinfo
  {journal} {Phys. Rev. A}\ }\textbf {\bibinfo {volume} {59}},  \bibinfo
  {pages} {230} (\bibinfo {year} {1999}{\natexlab{a}})}\BibitemShut {NoStop}%
\bibitem [{\citenamefont {Dzuba}\ \emph
  {et~al.}(1999{\natexlab{b}})\citenamefont {Dzuba}, \citenamefont {Flambaum},
  and \citenamefont {Webb}}]{DzuFlaWebPRL1999}%
  \BibitemOpen
  \bibfield  {author} {\bibinfo {author} {\bibfnamefont {V.~A.}\ \bibnamefont
  {Dzuba}}, \bibinfo {author} {\bibfnamefont {V.~V.}\ \bibnamefont {Flambaum}},
   and \bibinfo {author} {\bibfnamefont {J.~K.}\ \bibnamefont {Webb}},
  }\href {\doibase 10.1103/PhysRevLett.82.888} {\bibfield  {journal} {\bibinfo
  {journal} {Phys. Rev. Lett.}\ }\textbf {\bibinfo {volume} {82}},  \bibinfo
  {pages} {888} (\bibinfo {year} {1999}{\natexlab{b}})}\BibitemShut {NoStop}%
\bibitem [{\citenamefont {Lisdat}\ \emph {et~al.}(2016)\citenamefont {Lisdat},
  \citenamefont {Grosche}, \citenamefont {Quintin}, \citenamefont {Shi},
  \citenamefont {Raupach}, \citenamefont {Grebing}, \citenamefont {Nicolodi},
  \citenamefont {Stefani}, \citenamefont {Al-Masoudi}, \citenamefont
  {D{\"{o}}rscher}, \citenamefont {H{\"{a}}fner}, \citenamefont {Robyr},
  \citenamefont {Chiodo}, \citenamefont {Bilicki}, \citenamefont {Bookjans},
  \citenamefont {Koczwara}, \citenamefont {Koke}, \citenamefont {Kuhl},
  \citenamefont {Wiotte}, \citenamefont {Meynadier}, \citenamefont {Camisard},
  \citenamefont {Abgrall}, \citenamefont {Lours}, \citenamefont {Legero},
  \citenamefont {Schnatz}, \citenamefont {Sterr}, \citenamefont {Denker},
  \citenamefont {Chardonnet}, \citenamefont {{Le Coq}}, \citenamefont
  {Santarelli}, \citenamefont {Amy-Klein}, \citenamefont {{Le Targat}},
  \citenamefont {Lodewyck}, \citenamefont {Lopez}, and \citenamefont
  {Pottie}}]{Lisdat2016}%
  \BibitemOpen
  \bibfield  {author} {\bibinfo {author} {\bibfnamefont {C.}~\bibnamefont
  {Lisdat}}, \bibinfo {author} {\bibfnamefont {G.}~\bibnamefont {Grosche}},
  \bibinfo {author} {\bibfnamefont {N.}~\bibnamefont {Quintin}}, \bibinfo
  {author} {\bibfnamefont {C.}~\bibnamefont {Shi}}, \bibinfo {author}
  {\bibfnamefont {S.}~\bibnamefont {Raupach}}, \bibinfo {author} {\bibfnamefont
  {C.}~\bibnamefont {Grebing}}, \bibinfo {author} {\bibfnamefont
  {D.}~\bibnamefont {Nicolodi}}, \bibinfo {author} {\bibfnamefont
  {F.}~\bibnamefont {Stefani}}, \bibinfo {author} {\bibfnamefont
  {A.}~\bibnamefont {Al-Masoudi}}, \bibinfo {author} {\bibfnamefont
  {S.}~\bibnamefont {D{\"{o}}rscher}}, \bibinfo {author} {\bibfnamefont
  {S.}~\bibnamefont {H{\"{a}}fner}}, \bibinfo {author} {\bibfnamefont {J.-L.}
  \bibnamefont {Robyr}}, \bibinfo {author} {\bibfnamefont {N.}~\bibnamefont
  {Chiodo}}, \bibinfo {author} {\bibfnamefont {S.}~\bibnamefont {Bilicki}},
  \bibinfo {author} {\bibfnamefont {E.}~\bibnamefont {Bookjans}}, \bibinfo
  {author} {\bibfnamefont {A.}~\bibnamefont {Koczwara}}, \bibinfo {author}
  {\bibfnamefont {S.}~\bibnamefont {Koke}}, \bibinfo {author} {\bibfnamefont
  {A.}~\bibnamefont {Kuhl}}, \bibinfo {author} {\bibfnamefont {F.}~\bibnamefont
  {Wiotte}}, \bibinfo {author} {\bibfnamefont {F.}~\bibnamefont {Meynadier}},
  \bibinfo {author} {\bibfnamefont {E.}~\bibnamefont {Camisard}}, \bibinfo
  {author} {\bibfnamefont {M.}~\bibnamefont {Abgrall}}, \bibinfo {author}
  {\bibfnamefont {M.}~\bibnamefont {Lours}}, \bibinfo {author} {\bibfnamefont
  {T.}~\bibnamefont {Legero}}, \bibinfo {author} {\bibfnamefont
  {H.}~\bibnamefont {Schnatz}}, \bibinfo {author} {\bibfnamefont
  {U.}~\bibnamefont {Sterr}}, \bibinfo {author} {\bibfnamefont
  {H.}~\bibnamefont {Denker}}, \bibinfo {author} {\bibfnamefont
  {C.}~\bibnamefont {Chardonnet}}, \bibinfo {author} {\bibfnamefont
  {Y.}~\bibnamefont {{Le Coq}}}, \bibinfo {author} {\bibfnamefont
  {G.}~\bibnamefont {Santarelli}}, \bibinfo {author} {\bibfnamefont
  {A.}~\bibnamefont {Amy-Klein}}, \bibinfo {author} {\bibfnamefont
  {R.}~\bibnamefont {{Le Targat}}}, \bibinfo {author} {\bibfnamefont
  {J.}~\bibnamefont {Lodewyck}}, \bibinfo {author} {\bibfnamefont
  {O.}~\bibnamefont {Lopez}},  and \bibinfo {author} {\bibfnamefont {P.-E.}
  \bibnamefont {Pottie}},  }\href {\doibase 10.1038/ncomms12443} {\bibfield
  {journal} {\bibinfo  {journal} {Nat. Commun.}\ }\textbf {\bibinfo {volume}
  {7}},  \bibinfo {pages} {12443} (\bibinfo {year} {2016})}\BibitemShut
  {NoStop}%
\bibitem [{\citenamefont {King}\ \emph {et~al.}(2012)\citenamefont {King},
  \citenamefont {Godun}, \citenamefont {Webster}, \citenamefont {Margolis},
  \citenamefont {Johnson}, \citenamefont {Szymaniec}, \citenamefont {Baird},
  and \citenamefont {Gill}}]{King2012}%
  \BibitemOpen
  \bibfield  {author} {\bibinfo {author} {\bibfnamefont {S.~A.}\ \bibnamefont
  {King}}, \bibinfo {author} {\bibfnamefont {R.~M.}\ \bibnamefont {Godun}},
  \bibinfo {author} {\bibfnamefont {S.~A.}\ \bibnamefont {Webster}}, \bibinfo
  {author} {\bibfnamefont {H.~S.}\ \bibnamefont {Margolis}}, \bibinfo {author}
  {\bibfnamefont {L.~A.~M.}\ \bibnamefont {Johnson}}, \bibinfo {author}
  {\bibfnamefont {K.}~\bibnamefont {Szymaniec}}, \bibinfo {author}
  {\bibfnamefont {P.~E.~G.}\ \bibnamefont {Baird}},  and \bibinfo {author}
  {\bibfnamefont {P.}~\bibnamefont {Gill}},  }\href {\doibase
  10.1088/1367-2630/14/1/013045} {\bibfield  {journal} {\bibinfo  {journal}
  {New J. Phys.}\ }\textbf {\bibinfo {volume} {14}},  \bibinfo {pages} {013045}
  (\bibinfo {year} {2012})}\BibitemShut {NoStop}%
\bibitem [{\citenamefont {{Le Targat}}\ \emph {et~al.}(2013)\citenamefont {{Le
  Targat}}, \citenamefont {Lorini}, \citenamefont {{Le Coq}}, \citenamefont
  {Zawada}, \citenamefont {Gu{\'{e}}na}, \citenamefont {Abgrall}, \citenamefont
  {Gurov}, \citenamefont {Rosenbusch}, \citenamefont {Rovera}, \citenamefont
  {Nag{\'{o}}rny}, \citenamefont {Gartman}, \citenamefont {Westergaard},
  \citenamefont {Tobar}, \citenamefont {Lours}, \citenamefont {Santarelli},
  \citenamefont {Clairon}, \citenamefont {Bize}, \citenamefont {Laurent},
  \citenamefont {Lemonde}, and \citenamefont {Lodewyck}}]{LeTargat2013}%
  \BibitemOpen
  \bibfield  {author} {\bibinfo {author} {\bibfnamefont {R.}~\bibnamefont {{Le
  Targat}}}, \bibinfo {author} {\bibfnamefont {L.}~\bibnamefont {Lorini}},
  \bibinfo {author} {\bibfnamefont {Y.}~\bibnamefont {{Le Coq}}}, \bibinfo
  {author} {\bibfnamefont {M.}~\bibnamefont {Zawada}}, \bibinfo {author}
  {\bibfnamefont {J.}~\bibnamefont {Gu{\'{e}}na}}, \bibinfo {author}
  {\bibfnamefont {M.}~\bibnamefont {Abgrall}}, \bibinfo {author} {\bibfnamefont
  {M.}~\bibnamefont {Gurov}}, \bibinfo {author} {\bibfnamefont
  {P.}~\bibnamefont {Rosenbusch}}, \bibinfo {author} {\bibfnamefont {D.~G.}
  \bibnamefont {Rovera}}, \bibinfo {author} {\bibfnamefont {B.}~\bibnamefont
  {Nag{\'{o}}rny}}, \bibinfo {author} {\bibfnamefont {R.}~\bibnamefont
  {Gartman}}, \bibinfo {author} {\bibfnamefont {P.~G.}\ \bibnamefont
  {Westergaard}}, \bibinfo {author} {\bibfnamefont {M.~E.}\ \bibnamefont
  {Tobar}}, \bibinfo {author} {\bibfnamefont {M.}~\bibnamefont {Lours}},
  \bibinfo {author} {\bibfnamefont {G.}~\bibnamefont {Santarelli}}, \bibinfo
  {author} {\bibfnamefont {A.}~\bibnamefont {Clairon}}, \bibinfo {author}
  {\bibfnamefont {S.}~\bibnamefont {Bize}}, \bibinfo {author} {\bibfnamefont
  {P.}~\bibnamefont {Laurent}}, \bibinfo {author} {\bibfnamefont
  {P.}~\bibnamefont {Lemonde}},  and \bibinfo {author} {\bibfnamefont
  {J.}~\bibnamefont {Lodewyck}},  }\href {\doibase 10.1038/ncomms3109}
  {\bibfield  {journal} {\bibinfo  {journal} {Nat. Commun.}\ }\textbf {\bibinfo
  {volume} {4}},  \bibinfo {pages} {2109} (\bibinfo {year} {2013})}\BibitemShut
  {NoStop}%
\bibitem [{\citenamefont {Hill}\ \emph {et~al.}(2016)\citenamefont {Hill},
  \citenamefont {Hobson}, \citenamefont {Bowden}, \citenamefont {Bridge},
  \citenamefont {Donnellan}, \citenamefont {Curtis}, and \citenamefont
  {Gill}}]{Hill2016}%
  \BibitemOpen
  \bibfield  {author} {\bibinfo {author} {\bibfnamefont {I.~R.}\ \bibnamefont
  {Hill}}, \bibinfo {author} {\bibfnamefont {R.}~\bibnamefont {Hobson}},
  \bibinfo {author} {\bibfnamefont {W.}~\bibnamefont {Bowden}}, \bibinfo
  {author} {\bibfnamefont {E.~M.}\ \bibnamefont {Bridge}}, \bibinfo {author}
  {\bibfnamefont {S.}~\bibnamefont {Donnellan}}, \bibinfo {author}
  {\bibfnamefont {E.~A.}\ \bibnamefont {Curtis}},  and \bibinfo {author}
  {\bibfnamefont {P.}~\bibnamefont {Gill}},  }\href {\doibase
  10.1088/1742-6596/723/1/012019} {\bibfield  {journal} {\bibinfo  {journal}
  {J. Phys. Conf. Ser.}\ }\textbf {\bibinfo {volume} {723}},  \bibinfo {pages}
  {012019} (\bibinfo {year} {2016})}\BibitemShut {NoStop}%
\bibitem [{\citenamefont {Grebing}\ \emph {et~al.}(2016)\citenamefont
  {Grebing}, \citenamefont {Al-Masoudi}, \citenamefont {D{\"{o}}rscher},
  \citenamefont {H{\"{a}}fner}, \citenamefont {Gerginov}, \citenamefont
  {Weyers}, \citenamefont {Lipphardt}, \citenamefont {Riehle}, \citenamefont
  {Sterr}, and \citenamefont {Lisdat}}]{Grebing2016}%
  \BibitemOpen
  \bibfield  {author} {\bibinfo {author} {\bibfnamefont {C.}~\bibnamefont
  {Grebing}}, \bibinfo {author} {\bibfnamefont {A.}~\bibnamefont {Al-Masoudi}},
  \bibinfo {author} {\bibfnamefont {S.}~\bibnamefont {D{\"{o}}rscher}},
  \bibinfo {author} {\bibfnamefont {S.}~\bibnamefont {H{\"{a}}fner}}, \bibinfo
  {author} {\bibfnamefont {V.}~\bibnamefont {Gerginov}}, \bibinfo {author}
  {\bibfnamefont {S.}~\bibnamefont {Weyers}}, \bibinfo {author} {\bibfnamefont
  {B.}~\bibnamefont {Lipphardt}}, \bibinfo {author} {\bibfnamefont
  {F.}~\bibnamefont {Riehle}}, \bibinfo {author} {\bibfnamefont
  {U.}~\bibnamefont {Sterr}},  and \bibinfo {author} {\bibfnamefont
  {C.}~\bibnamefont {Lisdat}},  }\href {\doibase 10.1364/OPTICA.3.000563}
  {\bibfield  {journal} {\bibinfo  {journal} {Optica}\ }\textbf {\bibinfo
  {volume} {3}},  \bibinfo {pages} {563} (\bibinfo {year} {2016})}\BibitemShut
  {NoStop}%
\bibitem [{\citenamefont {Tyumenev}\ \emph {et~al.}(2016)\citenamefont
  {Tyumenev}, \citenamefont {Favier}, \citenamefont {Bilicki}, \citenamefont
  {Bookjans}, \citenamefont {Targat}, \citenamefont {Lodewyck}, \citenamefont
  {Nicolodi}, \citenamefont {Coq}, \citenamefont {Abgrall}, \citenamefont
  {Gu{\'{e}}na}, \citenamefont {Sarlo}, and \citenamefont
  {Bize}}]{Tyumenev2016}%
  \BibitemOpen
  \bibfield  {author} {\bibinfo {author} {\bibfnamefont {R.}~\bibnamefont
  {Tyumenev}}, \bibinfo {author} {\bibfnamefont {M.}~\bibnamefont {Favier}},
  \bibinfo {author} {\bibfnamefont {S.}~\bibnamefont {Bilicki}}, \bibinfo
  {author} {\bibfnamefont {E.}~\bibnamefont {Bookjans}}, \bibinfo {author}
  {\bibfnamefont {R.~L.}\ \bibnamefont {Targat}}, \bibinfo {author}
  {\bibfnamefont {J.}~\bibnamefont {Lodewyck}}, \bibinfo {author}
  {\bibfnamefont {D.}~\bibnamefont {Nicolodi}}, \bibinfo {author}
  {\bibfnamefont {Y.~L.}\ \bibnamefont {Coq}}, \bibinfo {author} {\bibfnamefont
  {M.}~\bibnamefont {Abgrall}}, \bibinfo {author} {\bibfnamefont
  {J.}~\bibnamefont {Gu{\'{e}}na}}, \bibinfo {author} {\bibfnamefont {L.~D.}
  \bibnamefont {Sarlo}},  and \bibinfo {author} {\bibfnamefont
  {S.}~\bibnamefont {Bize}},  }\href {\doibase 10.1088/1367-2630/18/11/113002}
  {\bibfield  {journal} {\bibinfo  {journal} {New J. Phys.}\ }\textbf {\bibinfo
  {volume} {18}},  \bibinfo {pages} {113002} (\bibinfo {year}
  {2016})}\BibitemShut {NoStop}%
\bibitem [{\citenamefont {Sanner}\ \emph {et~al.}(2019)\citenamefont {Sanner},
  \citenamefont {Huntemann}, \citenamefont {Lange}, \citenamefont {Tamm},
  \citenamefont {Peik}, \citenamefont {Safronova}, and \citenamefont
  {Porsev}}]{Sanner2019}%
  \BibitemOpen
  \bibfield  {author} {\bibinfo {author} {\bibfnamefont {C.}~\bibnamefont
  {Sanner}}, \bibinfo {author} {\bibfnamefont {N.}~\bibnamefont {Huntemann}},
  \bibinfo {author} {\bibfnamefont {R.}~\bibnamefont {Lange}}, \bibinfo
  {author} {\bibfnamefont {C.}~\bibnamefont {Tamm}}, \bibinfo {author}
  {\bibfnamefont {E.}~\bibnamefont {Peik}}, \bibinfo {author} {\bibfnamefont
  {M.~S.}\ \bibnamefont {Safronova}},  and \bibinfo {author} {\bibfnamefont
  {S.~G.}\ \bibnamefont {Porsev}},  }\href {\doibase 10.1038/s41586-019-0972-2}
  {\bibfield  {journal} {\bibinfo  {journal} {Nature}\ }\textbf {\bibinfo
  {volume} {567}},  \bibinfo {pages} {204} (\bibinfo {year} {2019})},  \Eprint
  {http://arxiv.org/abs/1809.10742v1} {arXiv:1809.10742v1} \BibitemShut
  {NoStop}%
\bibitem [{\citenamefont {Delva}\ \emph {et~al.}(2017)\citenamefont {Delva},
  \citenamefont {Lodewyck}, \citenamefont {Bilicki}, \citenamefont {Bookjans},
  \citenamefont {Vallet}, \citenamefont {{Le Targat}}, \citenamefont {Pottie},
  \citenamefont {Guerlin}, \citenamefont {Meynadier}, \citenamefont {{Le
  Poncin-Lafitte}}, \citenamefont {Lopez}, \citenamefont {Amy-Klein},
  \citenamefont {Lee}, \citenamefont {Quintin}, \citenamefont {Lisdat},
  \citenamefont {Al-Masoudi}, \citenamefont {Dorscher}, \citenamefont
  {Grebing}, \citenamefont {Grosche}, \citenamefont {Kuhl}, \citenamefont
  {Raupach}, \citenamefont {Sterr}, \citenamefont {Hill}, \citenamefont
  {Hobson}, \citenamefont {Bowden}, \citenamefont {Kronjager}, \citenamefont
  {Marra}, \citenamefont {Rolland}, \citenamefont {Baynes}, \citenamefont
  {Margolis}, and \citenamefont {Gill}}]{Delva2017}%
  \BibitemOpen
  \bibfield  {author} {\bibinfo {author} {\bibfnamefont {P.}~\bibnamefont
  {Delva}}, \bibinfo {author} {\bibfnamefont {J.}~\bibnamefont {Lodewyck}},
  \bibinfo {author} {\bibfnamefont {S.}~\bibnamefont {Bilicki}}, \bibinfo
  {author} {\bibfnamefont {E.}~\bibnamefont {Bookjans}}, \bibinfo {author}
  {\bibfnamefont {G.}~\bibnamefont {Vallet}}, \bibinfo {author} {\bibfnamefont
  {R.}~\bibnamefont {{Le Targat}}}, \bibinfo {author} {\bibfnamefont {P.-E.}
  \bibnamefont {Pottie}}, \bibinfo {author} {\bibfnamefont {C.}~\bibnamefont
  {Guerlin}}, \bibinfo {author} {\bibfnamefont {F.}~\bibnamefont {Meynadier}},
  \bibinfo {author} {\bibfnamefont {C.}~\bibnamefont {{Le Poncin-Lafitte}}},
  \bibinfo {author} {\bibfnamefont {O.}~\bibnamefont {Lopez}}, \bibinfo
  {author} {\bibfnamefont {A.}~\bibnamefont {Amy-Klein}}, \bibinfo {author}
  {\bibfnamefont {W.-K.}\ \bibnamefont {Lee}}, \bibinfo {author} {\bibfnamefont
  {N.}~\bibnamefont {Quintin}}, \bibinfo {author} {\bibfnamefont
  {C.}~\bibnamefont {Lisdat}}, \bibinfo {author} {\bibfnamefont
  {A.}~\bibnamefont {Al-Masoudi}}, \bibinfo {author} {\bibfnamefont
  {S.}~\bibnamefont {Dorscher}}, \bibinfo {author} {\bibfnamefont
  {C.}~\bibnamefont {Grebing}}, \bibinfo {author} {\bibfnamefont
  {G.}~\bibnamefont {Grosche}}, \bibinfo {author} {\bibfnamefont
  {A.}~\bibnamefont {Kuhl}}, \bibinfo {author} {\bibfnamefont {S.}~\bibnamefont
  {Raupach}}, \bibinfo {author} {\bibfnamefont {U.}~\bibnamefont {Sterr}},
  \bibinfo {author} {\bibfnamefont {I.~R.}\ \bibnamefont {Hill}}, \bibinfo
  {author} {\bibfnamefont {R.}~\bibnamefont {Hobson}}, \bibinfo {author}
  {\bibfnamefont {W.}~\bibnamefont {Bowden}}, \bibinfo {author} {\bibfnamefont
  {J.}~\bibnamefont {Kronjager}}, \bibinfo {author} {\bibfnamefont
  {G.}~\bibnamefont {Marra}}, \bibinfo {author} {\bibfnamefont
  {A.}~\bibnamefont {Rolland}}, \bibinfo {author} {\bibfnamefont {F.~N.}
  \bibnamefont {Baynes}}, \bibinfo {author} {\bibfnamefont {H.~S.}
  \bibnamefont {Margolis}},  and \bibinfo {author} {\bibfnamefont
  {P.}~\bibnamefont {Gill}},  }\href {\doibase 10.1103/PhysRevLett.118.221102}
  {\bibfield  {journal} {\bibinfo  {journal} {Phys. Rev. Lett.}\ }\textbf
  {\bibinfo {volume} {118}},  \bibinfo {pages} {221102} (\bibinfo {year}
  {2017})}\BibitemShut {NoStop}%
\bibitem [{\citenamefont {Angstmann}\ \emph {et~al.}(2004)\citenamefont
  {Angstmann}, \citenamefont {Dzuba}, and \citenamefont
  {Flambaum}}]{Angstmann2004}%
  \BibitemOpen
  \bibfield  {author} {\bibinfo {author} {\bibfnamefont {E.~J.}\ \bibnamefont
  {Angstmann}}, \bibinfo {author} {\bibfnamefont {V.~A.}\ \bibnamefont
  {Dzuba}},  and \bibinfo {author} {\bibfnamefont {V.~V.}\ \bibnamefont
  {Flambaum}},  }\href {\doibase 10.1103/PhysRevA.70.014102} {\bibfield
  {journal} {\bibinfo  {journal} {Phys. Rev. A}\ }\textbf {\bibinfo {volume}
  {70}},  \bibinfo {pages} {014102} (\bibinfo {year} {2004})}\BibitemShut
  {NoStop}%
\bibitem [{\citenamefont {Dzuba} and \citenamefont
  {Flambaum}(2008)}]{DzubaKrel2008}%
  \BibitemOpen
  \bibfield  {author} {\bibinfo {author} {\bibfnamefont {V.~A.}\ \bibnamefont
  {Dzuba}} and \bibinfo {author} {\bibfnamefont {V.~V.}\ \bibnamefont
  {Flambaum}},  }\href {\doibase 10.1103/PhysRevA.77.012515} {\bibfield
  {journal} {\bibinfo  {journal} {Phys. Rev. A}\ }\textbf {\bibinfo {volume}
  {77}},  \bibinfo {pages} {012515} (\bibinfo {year} {2008})}\BibitemShut
  {NoStop}%
\bibitem [{\citenamefont {Freese}\ \emph {et~al.}(2013)\citenamefont {Freese},
  \citenamefont {Lisanti}, and \citenamefont {Savage}}]{Freese2013}%
  \BibitemOpen
  \bibfield  {author} {\bibinfo {author} {\bibfnamefont {K.}~\bibnamefont
  {Freese}}, \bibinfo {author} {\bibfnamefont {M.}~\bibnamefont {Lisanti}}, 
  and \bibinfo {author} {\bibfnamefont {C.}~\bibnamefont {Savage}},  }\href
  {\doibase 10.1103/RevModPhys.85.1561} {\bibfield  {journal} {\bibinfo
  {journal} {Rev. Mod. Phys.}\ }\textbf {\bibinfo {volume} {85}},  \bibinfo
  {pages} {1561} (\bibinfo {year} {2013})}\BibitemShut {NoStop}%
\bibitem [{\citenamefont {Roberts}\ \emph {et~al.}(2017)\citenamefont
  {Roberts}, \citenamefont {Blewitt}, \citenamefont {Dailey}, \citenamefont
  {Murphy}, \citenamefont {Pospelov}, \citenamefont {Rollings}, \citenamefont
  {Sherman}, \citenamefont {Williams}, and \citenamefont
  {Derevianko}}]{GPSDM2017}%
  \BibitemOpen
  \bibfield  {author} {\bibinfo {author} {\bibfnamefont {B.~M.}\ \bibnamefont
  {Roberts}}, \bibinfo {author} {\bibfnamefont {G.}~\bibnamefont {Blewitt}},
  \bibinfo {author} {\bibfnamefont {C.}~\bibnamefont {Dailey}}, \bibinfo
  {author} {\bibfnamefont {M.}~\bibnamefont {Murphy}}, \bibinfo {author}
  {\bibfnamefont {M.}~\bibnamefont {Pospelov}}, \bibinfo {author}
  {\bibfnamefont {A.}~\bibnamefont {Rollings}}, \bibinfo {author}
  {\bibfnamefont {J.}~\bibnamefont {Sherman}}, \bibinfo {author} {\bibfnamefont
  {W.}~\bibnamefont {Williams}},  and \bibinfo {author} {\bibfnamefont
  {A.}~\bibnamefont {Derevianko}},  }\href {\doibase
  10.1038/s41467-017-01440-4} {\bibfield  {journal} {\bibinfo  {journal} {Nat.
  Commun.}\ }\textbf {\bibinfo {volume} {8}},  \bibinfo {pages} {1195}
  (\bibinfo {year} {2017})},  \Eprint {http://arxiv.org/abs/1704.06844}
  {arXiv:1704.06844} \BibitemShut {NoStop}%
\bibitem [{\citenamefont {Roberts}\ \emph {et~al.}(2018)\citenamefont
  {Roberts}, \citenamefont {Blewitt}, \citenamefont {Dailey}, and
  \citenamefont {Derevianko}}]{GPSDM2018}%
  \BibitemOpen
  \bibfield  {author} {\bibinfo {author} {\bibfnamefont {B.~M.}\ \bibnamefont
  {Roberts}}, \bibinfo {author} {\bibfnamefont {G.}~\bibnamefont {Blewitt}},
  \bibinfo {author} {\bibfnamefont {C.}~\bibnamefont {Dailey}},  and \bibinfo
  {author} {\bibfnamefont {A.}~\bibnamefont {Derevianko}},  }\href {\doibase
  10.1103/PhysRevD.97.083009} {\bibfield  {journal} {\bibinfo  {journal} {Phys.
  Rev. D}\ }\textbf {\bibinfo {volume} {97}},  \bibinfo {pages} {083009}
  (\bibinfo {year} {2018})},  \Eprint {http://arxiv.org/abs/1803.10264}
  {arXiv:1803.10264} \BibitemShut {NoStop}%
\bibitem [{\citenamefont {Wcis{\l}o}\ \emph {et~al.}(2016)\citenamefont
  {Wcis{\l}o}, \citenamefont {Morzy{\'{n}}ski}, \citenamefont {Bober},
  \citenamefont {Cygan}, \citenamefont {Lisak}, \citenamefont {Ciury{\l}o},
  and \citenamefont {Zawada}}]{Wcislo2016}%
  \BibitemOpen
  \bibfield  {author} {\bibinfo {author} {\bibfnamefont {P.}~\bibnamefont
  {Wcis{\l}o}}, \bibinfo {author} {\bibfnamefont {P.}~\bibnamefont
  {Morzy{\'{n}}ski}}, \bibinfo {author} {\bibfnamefont {M.}~\bibnamefont
  {Bober}}, \bibinfo {author} {\bibfnamefont {A.}~\bibnamefont {Cygan}},
  \bibinfo {author} {\bibfnamefont {D.}~\bibnamefont {Lisak}}, \bibinfo
  {author} {\bibfnamefont {R.}~\bibnamefont {Ciury{\l}o}},  and \bibinfo
  {author} {\bibfnamefont {M.}~\bibnamefont {Zawada}},  }\href {\doibase
  10.1038/s41550-016-0009} {\bibfield  {journal} {\bibinfo  {journal} {Nat.
  Astron.}\ }\textbf {\bibinfo {volume} {1}},  \bibinfo {pages} {0009}
  (\bibinfo {year} {2016})}\BibitemShut {NoStop}%
\bibitem [{\citenamefont {Wcis{\l}o}\ \emph {et~al.}(2018)\citenamefont
  {Wcis{\l}o}, \citenamefont {Ablewski}, \citenamefont {Beloy}, \citenamefont
  {Bilicki}, \citenamefont {Bober}, \citenamefont {Brown}, \citenamefont
  {Fasano}, \citenamefont {Ciury{\l}o}, \citenamefont {Hachisu}, \citenamefont
  {Ido}, \citenamefont {Lodewyck}, \citenamefont {Ludlow}, \citenamefont
  {McGrew}, \citenamefont {Morzy{\'{n}}ski}, \citenamefont {Nicolodi},
  \citenamefont {Schioppo}, \citenamefont {Sekido}, \citenamefont {{Le
  Targat}}, \citenamefont {Wolf}, \citenamefont {Zhang}, \citenamefont
  {Zjawin}, and \citenamefont {Zawada}}]{Wciso2018}%
  \BibitemOpen
  \bibfield  {author} {\bibinfo {author} {\bibfnamefont {P.}~\bibnamefont
  {Wcis{\l}o}}, \bibinfo {author} {\bibfnamefont {P.}~\bibnamefont {Ablewski}},
  \bibinfo {author} {\bibfnamefont {K.}~\bibnamefont {Beloy}}, \bibinfo
  {author} {\bibfnamefont {S.}~\bibnamefont {Bilicki}}, \bibinfo {author}
  {\bibfnamefont {M.}~\bibnamefont {Bober}}, \bibinfo {author} {\bibfnamefont
  {R.}~\bibnamefont {Brown}}, \bibinfo {author} {\bibfnamefont
  {R.}~\bibnamefont {Fasano}}, \bibinfo {author} {\bibfnamefont
  {R.}~\bibnamefont {Ciury{\l}o}}, \bibinfo {author} {\bibfnamefont
  {H.}~\bibnamefont {Hachisu}}, \bibinfo {author} {\bibfnamefont
  {T.}~\bibnamefont {Ido}}, \bibinfo {author} {\bibfnamefont {J.}~\bibnamefont
  {Lodewyck}}, \bibinfo {author} {\bibfnamefont {A.~D.}\ \bibnamefont
  {Ludlow}}, \bibinfo {author} {\bibfnamefont {W.~F.}\ \bibnamefont {McGrew}},
  \bibinfo {author} {\bibfnamefont {P.}~\bibnamefont {Morzy{\'{n}}ski}},
  \bibinfo {author} {\bibfnamefont {D.}~\bibnamefont {Nicolodi}}, \bibinfo
  {author} {\bibfnamefont {M.}~\bibnamefont {Schioppo}}, \bibinfo {author}
  {\bibfnamefont {M.}~\bibnamefont {Sekido}}, \bibinfo {author} {\bibfnamefont
  {R.}~\bibnamefont {{Le Targat}}}, \bibinfo {author} {\bibfnamefont
  {P.}~\bibnamefont {Wolf}}, \bibinfo {author} {\bibfnamefont {X.}~\bibnamefont
  {Zhang}}, \bibinfo {author} {\bibfnamefont {B.}~\bibnamefont {Zjawin}}, 
  and \bibinfo {author} {\bibfnamefont {M.}~\bibnamefont {Zawada}},  }\href
  {\doibase 10.1126/sciadv.aau4869} {\bibfield  {journal} {\bibinfo  {journal}
  {Sci. Adv.}\ }\textbf {\bibinfo {volume} {4}},  \bibinfo {pages} {eaau4869}
  (\bibinfo {year} {2018})},  \Eprint {http://arxiv.org/abs/1806.04762}
  {arXiv:1806.04762} \BibitemShut {NoStop}%
\bibitem [{\citenamefont {Bovy} and \citenamefont
  {Tremaine}(2012)}]{Bovy:2012tw}%
  \BibitemOpen
  \bibfield  {author} {\bibinfo {author} {\bibfnamefont {J.}~\bibnamefont
  {Bovy}} and \bibinfo {author} {\bibfnamefont {S.}~\bibnamefont
  {Tremaine}},  }\href {\doibase 10.1088/0004-637X/756/1/89} {\bibfield
  {journal} {\bibinfo  {journal} {Astrophys. J.}\ }\textbf {\bibinfo {volume}
  {756}},  \bibinfo {pages} {89} (\bibinfo {year} {2012})},  \Eprint
  {http://arxiv.org/abs/1205.4033} {arXiv:1205.4033} \BibitemShut {NoStop}%
\bibitem [{Note1()}]{Note1}%
  \BibitemOpen
  \bibinfo {note} {For definitiveness, and to be consistent with recent works,
  we take $\rho _{\protect \rm DM} =0.4\protect \ensuremath {\protect \tmspace
  +\thinmuskip {.1667em}{\protect \rm {GeV}}}\protect \ensuremath {\protect
  \tmspace +\thinmuskip {.1667em}{\protect \rm {cm}}}^{-3}$.}\BibitemShut
  {Stop}%
\bibitem [{\citenamefont {Raffelt}(1999)}]{Raffelt1999}%
  \BibitemOpen
  \bibfield  {author} {\bibinfo {author} {\bibfnamefont {G.~G.}\ \bibnamefont
  {Raffelt}},  }\href {\doibase 10.1146/annurev.nucl.49.1.163} {\bibfield
  {journal} {\bibinfo  {journal} {Annu. Rev. Nucl. Part. Sci.}\ }\textbf
  {\bibinfo {volume} {49}},  \bibinfo {pages} {163} (\bibinfo {year}
  {1999})}\BibitemShut {NoStop}%
\bibitem [{\citenamefont {Hirata}\ \emph {et~al.}(1988)\citenamefont {Hirata},
  \citenamefont {Kajita}, \citenamefont {Koshiba}, \citenamefont {Nakahata},
  \citenamefont {Oyama}, \citenamefont {Sato}, \citenamefont {Suzuki},
  \citenamefont {Takita}, \citenamefont {Totsuka}, \citenamefont {Kifune},
  \citenamefont {Suda}, \citenamefont {Takahashi}, \citenamefont {Tanimori},
  \citenamefont {Miyano}, \citenamefont {Yamada}, \citenamefont {Beier},
  \citenamefont {Feldscher}, \citenamefont {Frati}, \citenamefont {Kim},
  \citenamefont {Mann}, \citenamefont {Newcomer}, \citenamefont {{Van Berg}},
  \citenamefont {Zhang}, and \citenamefont {Cortez}}]{Hirata1988}%
  \BibitemOpen
  \bibfield  {author} {\bibinfo {author} {\bibfnamefont {K.~S.}\ \bibnamefont
  {Hirata}}, \bibinfo {author} {\bibfnamefont {T.}~\bibnamefont {Kajita}},
  \bibinfo {author} {\bibfnamefont {M.}~\bibnamefont {Koshiba}}, \bibinfo
  {author} {\bibfnamefont {M.}~\bibnamefont {Nakahata}}, \bibinfo {author}
  {\bibfnamefont {Y.}~\bibnamefont {Oyama}}, \bibinfo {author} {\bibfnamefont
  {N.}~\bibnamefont {Sato}}, \bibinfo {author} {\bibfnamefont {A.}~\bibnamefont
  {Suzuki}}, \bibinfo {author} {\bibfnamefont {M.}~\bibnamefont {Takita}},
  \bibinfo {author} {\bibfnamefont {Y.}~\bibnamefont {Totsuka}}, \bibinfo
  {author} {\bibfnamefont {T.}~\bibnamefont {Kifune}}, \bibinfo {author}
  {\bibfnamefont {T.}~\bibnamefont {Suda}}, \bibinfo {author} {\bibfnamefont
  {K.}~\bibnamefont {Takahashi}}, \bibinfo {author} {\bibfnamefont
  {T.}~\bibnamefont {Tanimori}}, \bibinfo {author} {\bibfnamefont
  {K.}~\bibnamefont {Miyano}}, \bibinfo {author} {\bibfnamefont
  {M.}~\bibnamefont {Yamada}}, \bibinfo {author} {\bibfnamefont {E.~W.}
  \bibnamefont {Beier}}, \bibinfo {author} {\bibfnamefont {L.~R.}\ \bibnamefont
  {Feldscher}}, \bibinfo {author} {\bibfnamefont {W.}~\bibnamefont {Frati}},
  \bibinfo {author} {\bibfnamefont {S.~B.}\ \bibnamefont {Kim}}, \bibinfo
  {author} {\bibfnamefont {A.~K.}\ \bibnamefont {Mann}}, \bibinfo {author}
  {\bibfnamefont {F.~M.}\ \bibnamefont {Newcomer}}, \bibinfo {author}
  {\bibfnamefont {R.}~\bibnamefont {{Van Berg}}}, \bibinfo {author}
  {\bibfnamefont {W.}~\bibnamefont {Zhang}},  and \bibinfo {author}
  {\bibfnamefont {B.~G.}\ \bibnamefont {Cortez}},  }\href {\doibase
  10.1103/PhysRevD.38.448} {\bibfield  {journal} {\bibinfo  {journal} {Phys.
  Rev. D}\ }\textbf {\bibinfo {volume} {38}},  \bibinfo {pages} {448} (\bibinfo
  {year} {1988})}\BibitemShut {NoStop}%
\bibitem [{\citenamefont {Roberts} and \citenamefont
  {Derevianko}(2018)}]{RobertsAsymm2018}%
  \BibitemOpen
  \bibfield  {author} {\bibinfo {author} {\bibfnamefont {B.~M.}\ \bibnamefont
  {Roberts}} and \bibinfo {author} {\bibfnamefont {A.}~\bibnamefont
  {Derevianko}},  }\href {http://arxiv.org/abs/1803.00617} {   (\bibinfo {year}
  {2018})},  \Eprint {http://arxiv.org/abs/1803.00617} {arXiv:1803.00617}
  \BibitemShut {NoStop}%
\bibitem [{\citenamefont {Savalle}\ \emph {et~al.}(2019)\citenamefont
  {Savalle}, \citenamefont {Roberts}, \citenamefont {Frank}, \citenamefont
  {Pottie}, \citenamefont {McAllister}, \citenamefont {Dailey}, \citenamefont
  {Derevianko}, and \citenamefont {Wolf}}]{Savalle2019}%
  \BibitemOpen
  \bibfield  {author} {\bibinfo {author} {\bibfnamefont {E.}~\bibnamefont
  {Savalle}}, \bibinfo {author} {\bibfnamefont {B.~M.}\ \bibnamefont
  {Roberts}}, \bibinfo {author} {\bibfnamefont {F.}~\bibnamefont {Frank}},
  \bibinfo {author} {\bibfnamefont {P.-E.}\ \bibnamefont {Pottie}}, \bibinfo
  {author} {\bibfnamefont {B.~T.}\ \bibnamefont {McAllister}}, \bibinfo
  {author} {\bibfnamefont {C.}~\bibnamefont {Dailey}}, \bibinfo {author}
  {\bibfnamefont {A.}~\bibnamefont {Derevianko}},  and \bibinfo {author}
  {\bibfnamefont {P.}~\bibnamefont {Wolf}},  }\href
  {https://arxiv.org/abs/1902.07192} {   (\bibinfo {year} {2019})},  \Eprint
  {http://arxiv.org/abs/1902.07192} {arXiv:1902.07192} \BibitemShut {NoStop}%
\bibitem [{\citenamefont {Gregory}(2005)}]{GregoryBayesian2005}%
  \BibitemOpen
  \bibfield  {author} {\bibinfo {author} {\bibfnamefont {P.~C.}\ \bibnamefont
  {Gregory}}, }{\emph {\bibinfo
  {title} {Bayesian Logical Data Analysis for the Physical Sciences}}}\ (\bibinfo  {publisher}
  {Cambridge University Press},  \bibinfo {year} {2005})\BibitemShut {NoStop}%
\end{thebibliography}%

\end{document}